\documentclass[preprint,aps,prd,showpacs,nofootinbib]{revtex4-1}
\usepackage{epsfig}
\usepackage{amsmath,graphicx}
\usepackage{amsmath}
\usepackage{graphicx}
\usepackage{amsfonts}
\usepackage{amssymb}%
\providecommand{\U}[1]{\protect\rule{.1in}{.1in}}
\begin{document}

\title{Photon-pion transition form factor: BABAR puzzle is cracked }
\author{A. E. Dorokhov}
\affiliation{Joint Institute for Nuclear Research, Bogoliubov Laboratory of Theoretical
Physics, 141980 Dubna, Moscow region, Russian Federation;\\
Institute for Theoretical Problems of Microphysics, Moscow State University,
RU-119899, Moscow, Russian Federation}

\date{\today }

\begin{abstract}
Recently, the BABAR collaboration published (arXiv:0905.4778) data for the
photon-pion transition form factor $F_{\pi\gamma\gamma^{\ast}}\left(
Q^{2}\right)  $, which are in strong contradiction to the predictions of the
standard factorization approach to perturbative QCD. Immediately afterwards,
two mechanisms were suggested (A.E. Dorokhov, arXiv:0905.4577; A.V.
Radyushkin, arXiv:0906.0323), that logarithmically enhance the form factor
asymptotics and therefore provide a qualitatively satisfactory description of
the BABAR data. However, the physics of the BABAR effect was not fully
clarified. In the present work, based on a nonperturbative approach to the QCD
vacuum and on rather universal assumptions, we show that there exists two
asymptotic regimes for the pion transition form factor. One regime with
asymptotics $F_{\pi\gamma^{\ast}\gamma}\left(  Q^{2}\right)  \sim1/Q^{2}$
corresponds to the result of the standard QCD factorization approach, while
other violates the standard factorization and leads to asymptotic behavior as
$F_{\pi\gamma^{\ast}\gamma}\left(  Q^{2}\right)  \sim\ln\left(  Q^{2}\right)
/Q^{2}$. Furthermore, considering specific nonlocal chiral quark models, we
find the region of parameters, where the existing CELLO, CLEO and BABAR data
for the pion transition form factor are successfully described.

\end{abstract}
\maketitle

\section{Introduction}

In the years 1977-1981, the theory of hard exclusive processes was formulated
within the factorization approach to perturbative quantum chromodynamics
(pQCD)
\cite{Radyushkin:1977gp,Lepage:1979zb,Efremov:1978rn,Efremov:1979qk,Efremov:1978fi,Lepage:1980fj,Brodsky:1981rp}%
. The main ingredients of this approach are the operator product expansion
(OPE), the factorization theorems, and the pQCD evolution equations. In this
context, the form factor for the photon-pion transition $\gamma^{\ast}%
\gamma^{\ast}\rightarrow\pi^{0},$ with both photons being spacelike (with
photon virtualities $Q_{1}^{2},Q_{2}^{2}>0$), was considered in
\cite{Lepage:1980fj,Brodsky:1981rp}. Since only one hadron is involved, the
corresponding form factor $F_{\pi\gamma^{\ast}\gamma^{\ast}}(Q_{1}^{2}%
,Q_{2}^{2})$ has the simplest structure for the pQCD analysis among the hard
exclusive processes. The nonperturbative information about the pion is
accumulated in the pion distribution amplitude (DA) $\varphi_{\pi}\left(
x\right)  $ for the fraction $x$ of the longitudinal pion momenta $p$, carried
by a quark. Another simplification is, that the short-distance amplitude for
the $\gamma^{\ast}\gamma^{\ast}\rightarrow\pi^{0}$ transition is, to leading
order, just given by a single quark propagator. Finally, the photon-pion form
factor is related to the axial anomaly \cite{Adler:1969gk,Bell:1969ts}, when
both photons are real.

\begin{figure}[h]
\hspace*{-10mm}\includegraphics[width=0.8\textwidth]{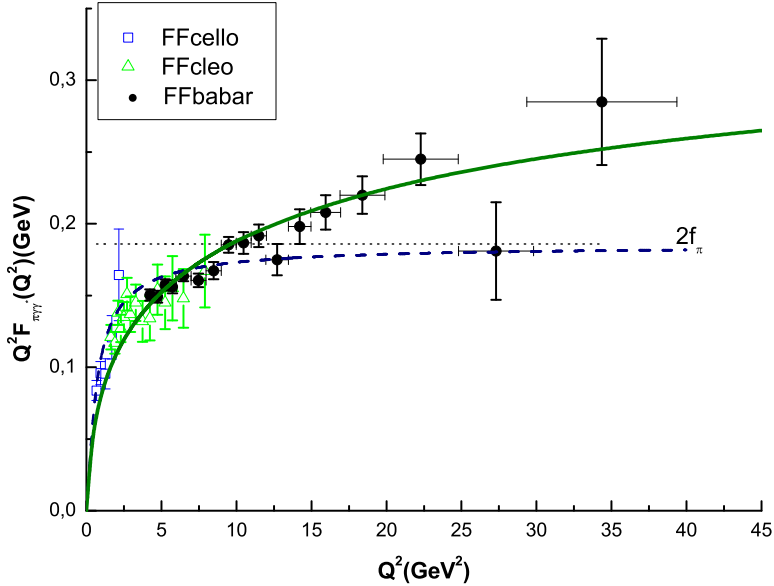}
\vspace*{-10mm}\caption{{\protect\footnotesize The transition form factor
$\gamma^{\ast}\gamma\rightarrow\pi^{0}$. The data are from the CELLO
\cite{Behrend:1990sr} (empty squares), CLEO \cite{Gronberg:1997fj} (empty
triangles) and BABAR (filled circles) \cite{:2009mc} Collaborations. The solid
line is the model of this work, the dashed line is the Brodsky-Lepage
prediction (\ref{FpqcdBL}), the short-dashed line is massless QCD asymptotic
limit ( \ref{FpqcdAS}).} }%
\label{Data}%
\end{figure}

\begin{figure}[h]
\hspace*{-10mm}\includegraphics[width=0.8\textwidth]{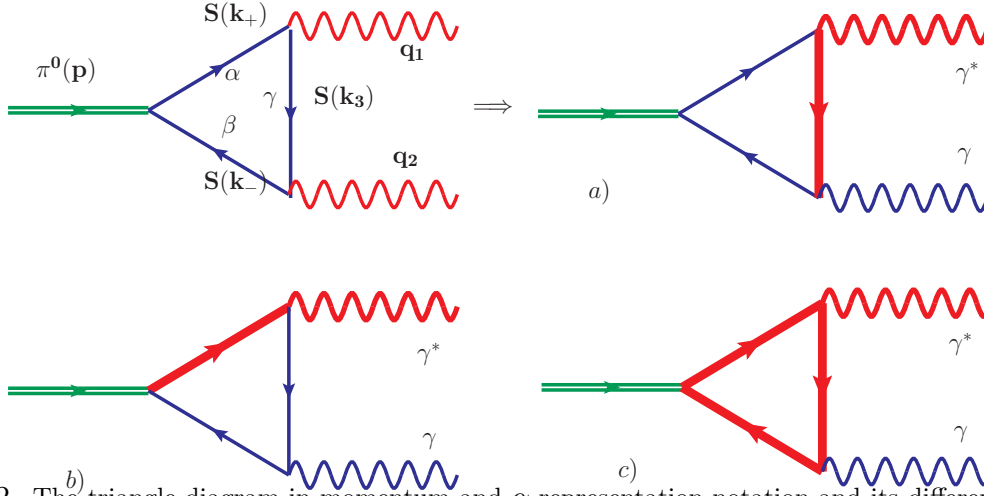}
\vspace*{-10mm}\caption{{\protect\footnotesize The triangle diagram in
momentum and }$\alpha${\protect\footnotesize -representation notation and its
different hard regimes for asymmetric kinematic. Hard photon is fat wavy line,
real photon is thin wavy line; the fat line is the hard propagator, the thin
line is the soft propagator: a) the standard factorization regime, b) the
regime violating standard factorization, c) the double logarithmic regime with
constant quark masses.}}%
\label{TriangleHard}%
\end{figure}

Experimentally, the easiest situation is, when one photon virtuality is small
and the other large. Under these conditions, the form factor $F_{\pi
\gamma^{\ast}\gamma}(Q^{2},0)$ was measured at $e^{+}e^{-}$ colliders by CELLO
\cite{Behrend:1990sr}, CLEO \cite{Gronberg:1997fj} Collaborations (Fig.
\ref{Data}). In the region of large virtualities $Q^{2}>>1$ GeV$^{2}$, the
pQCD factorization approach for exclusive processes predicts to leading order
in the strong coupling constant \cite{Lepage:1980fj,Brodsky:1981rp}
\begin{equation}
F_{\pi\gamma^{\ast}\gamma}^{\mathrm{pQCD}}(Q^{2},0)=\frac{2f_{\pi}}{3Q^{2}%
}J,\label{Fpqcd}%
\end{equation}
where%
\begin{equation}
J=\int_{0}^{1}dx\frac{\varphi_{\pi}\left(  x\right)  }{x}\label{J}%
\end{equation}
is the inverse moment of the pion DA, and $f_{\pi}=92.4$ MeV. The factor
$1/Q^{2}$ reflects the asymptotic property of the quark propagator connecting
two quark-photon vertices (Figs. \ref{TriangleHard}a and \ref{FactBL}). The
formula (\ref{Fpqcd}) is derived under the assumption, that the QCD dynamics
at large distances (the factor $Jf_{\pi}$) and the QCD dynamics at small
distances (the factor $1/Q^{2}$) is factorized. Moreover, under this
assumption, the asymptotics is reached already at the typical hadronic scale
of a few GeV$^{2}$. The pion DA $\varphi_{\pi}\left(  x\right)  $, in
addition, evolves in shape with the change of the renormalization scale
\cite{Efremov:1979qk,Lepage:1980fj} and asymptotically equals
\cite{Efremov:1978rn} $\varphi_{\pi}^{\mathrm{As}}\left(  x\right)  =6x\left(
1-x\right)  $. From this follows the famous asymptotic prediction (the
short-dashed line in Fig. \ref{Data})%
\begin{equation}
F_{\pi\gamma^{\ast}\gamma}^{\mathrm{pQCD,As}}(Q^{2},0)=\frac{2f_{\pi}}{Q^{2}%
}.\label{FpqcdAS}%
\end{equation}
To describe the soft nonperturbative region of $Q^{2}$, a simple interpolation
between the $Q^{2}\rightarrow0$ and $Q^{2}\rightarrow\infty$ limits has been
proposed by Brodsky and Lepage (BL) (the dashed line in Fig. \ref{Data})%
\begin{equation}
F_{\pi\gamma^{\ast}\gamma}^{\mathrm{BL}}(Q^{2},0)=\frac{1}{4\pi^{2}f_{\pi}%
}\frac{1}{1+Q^{2}/\left(  8\pi^{2}f_{\pi}^{2}\right)  }.\label{FpqcdBL}%
\end{equation}

Recently, the BABAR collaboration published new data (Fig. \ref{Data}) for the
$\gamma\gamma^{\ast}\rightarrow\pi^{0}$ transition form factor in the momentum
transfer range from 4 to 40 GeV$^{2}$ \cite{:2009mc}. They found the following
puzzling result: At $Q^{2}>10$ GeV$^{2}$ the measured form factor multiplied
by the photon virtuality $Q^{2}F_{\pi\gamma^{\ast}\gamma}(Q^{2},0)$ exceeds
the predicted asymptotic limit (\ref{FpqcdAS}) and, moreover, continues to
grow with increasing $Q^{2}$. This result is in strong contradiction to the
predictions of the standard QCD factorization approach mentioned above. The
BABAR data very well match the older data obtained by the CLEO collaboration
in the smaller $Q^{2}$ region, but extend to a much lager $Q^{2}$ values.
There is numerous literature discussing the BABAR effect. We refer here only
to the first two publications \cite{Dorokhov:2009dg,Radyushkin:2009zg}
appeared soon after the data were announced and it is these works that are the
most relevant for the following consideration. In these works two scenarios
were suggested, that logarithmically enhance the form factor asymptotics and
well describe the BABAR data. 

The first scenario \cite{Dorokhov:2009dg} uses the simple constituent quark
model \cite{Ametller:1983ec}. Within this model, the pion transition form
factor, determined by the quark-loop (triangle) diagram with a
momentum-independent quark mass $M_{q}$, is given by
\begin{equation}
F_{\pi\gamma\gamma^{\ast}}(Q^{2},0)=\frac{1}{4\pi^{2}f_{\pi}}\frac{m_{\pi}%
^{2}}{m_{\pi}^{2}+Q^{2}}\frac{1}{2\arcsin^{2}(\frac{m_{\pi}}{2M_{q}}%
)}\{2\arcsin^{2}(\frac{m_{\pi}}{2M_{q}})+\frac{1}{2}\ln^{2}\frac{\beta_{q}%
+1}{\beta_{q}-1}\},\label{Famet}%
\end{equation}
where $\beta_{q}=\sqrt{1+4M_{q}^{2}/Q^{2}}$. The form factor (\ref{Famet}) has
correct normalization at zero photon virtualities, by the axial anomaly, and
has double logarithmic asymptotics $\ln^{2}(Q^{2}/M_{q}^{2})/Q^{2}$ at large
$Q^{2}$. This asymptotics corresponds to the case when large virtuality pass
through all three quark propagators (Fig. \ref{TriangleHard}c). In
\cite{Dorokhov:2009dg} it was shown that the pion transition form factor
calculated from (\ref{Famet}) with the parameter $M_{q}=135$ MeV well
reproduces the BABAR data. However, this model has serious shortcomings.
Firstly, it has an incorrect chiral limit as $M_{q}\rightarrow0$ and $m_{\pi
}\rightarrow0$. Secondly, the corresponding integral for the decay constant
$f_{\pi}$ within this quark model is divergent and thus the model should be
regularized for consistency. After regularization, however, the double
logarithmic asymptotics is lost. Thirdly, just like in the Nambu--Jona-Lasinio
model, it uses a local $\gamma_{5}$ vertex for the quark-pion vertex and the
local quark propagator at all quark virtualities, in contradiction with pQCD,
where there is no $\gamma_{5}$ operator, no pion as a bound state and no
constituent quark mass. It is also well known, that in the local quark model
the distribution amplitude and distribution function of the pion are constants
\cite{Davidson:1994uv,RuizArriola:2002bp,Dorokhov:2000gu,Anikin:1999cx}.

Such flat (almost constant) pion DA was used in \cite{Radyushkin:2009zg} in
the context of the explanation of the BABAR data. The photon-pion transition
form factor was calculated by using expression from \cite{Lepage:1980fj} and
incorporating a light-cone wave function $\Psi\left(  x,k_{\bot}\right)  $
that has rapid falloff with respect to the light-front energy combination
$k_{\bot}^{2}/x\left(  1-x\right)  $. Within this approach, for a Gaussian
shape of the light-cone wave function and assuming a flat pion DA
$\varphi_{\pi}\left(  x\right)  =1$, the pion transition form factor is given
by
\begin{equation}
F_{\pi\gamma\gamma^{\ast}}^{\mathrm{As}}(Q^{2},0)=\frac{2}{3}\frac{f_{\pi}%
}{Q^{2}}\int_{0}^{1}\frac{dx}{x}\left[  1-\exp{\left(  -\frac{xQ^{2}}%
{2\sigma(1-x)}\right)  }\right]  ,\label{Frad}%
\end{equation}
and has logarithmically enhanced asymptotic behavior $\sim\ln{\left(
Q^{2}/2\sigma\right)  /}Q^{2}$. In \cite{Dorokhov:2009zx} it was demonstrated,
that the descriptions of the BABAR data in the model (\ref{Frad}) and in the
model (\ref{Famet}) practically coincide, if $\sigma=0.48$ GeV$^{2}$. In
\cite{Radyushkin:2009zg} it was also noted, that the use of such a wave
function is numerically close to the leading-order pQCD expression for the
photon-pion transition form factor with a modified quark propagator and a flat
pion DA%
\begin{equation}
F_{\pi\gamma\gamma^{\ast}}^{\mathrm{As}}(Q^{2},0)=\frac{2}{3}f_{\pi}\int
_{0}^{1}\frac{dx}{xQ^{2}+M^{2}},\label{FradM}%
\end{equation}
giving logarithmically enhanced asymptotics $F_{\pi\gamma\gamma^{\ast}%
}^{\mathrm{As}}(Q^{2},0)\sim\ln\left(  1+Q^{2}/M^{2}\right)  /Q^{2}.$ With
$M^{2}=0.6$ GeV$^{2}$ the BABAR data are well fitted. Another very important
feature of the scenario \cite{Radyushkin:2009zg} is, that it was argumented,
that there is no pQCD evolution modifications in the shape of flat pion DA.

However, these approaches did not give an answer to the following serious
questions. First of all, both expressions (\ref{Frad}) and (\ref{FradM}) do
not describe the full form factor, but only the leading asymptotic part. They
have incorrect normalization at $Q^{2}=0$, and are valid only at large photon
virtuality $Q^{2}>>1$ GeV$^{2}$. Furthermore, the appearance of the parameter
$M$ in the asymptotic formula (\ref{FradM}) is not justified. Moreover, as it
was emphasized in \cite{Radyushkin:2009zg}, the expression (\ref{FradM})
generates an infinite tower of false higher twist corrections $\left(
M^{2}/xQ^{2}\right)  ^{n}$ in contradiction with OPE. It is well known
\cite{Musatov:1997pu}, that there are only two terms in the OPE for the
handbag diagram for the pion transition form factor: the twist-2 and the
twist-4 terms. Also, the relation of the parameter $\sigma$ in (\ref{Frad}) or
$M$ in (\ref{FradM}) to the fundamental QCD parameters and their values
remained unclear. The decay constant $f_{\pi}$ is external parameter in this
approach and is not calculated dynamically. Finally, the origin of the flat DA
is not well understood. Most of the QCD sum rule and the instanton model
calculations lead to the endpoint suppressed amplitudes (see, e.g.
\cite{Bakulev:2002uc,Dorokhov:2002iu}). Below we show, how to generalize the
results (\ref{Famet})-(\ref{FradM}), and how to avoid the above mentioned
problems with the interpretation of the BABAR data.

There are several QCD based approaches to treat the nonperturbative aspects of
strong interactions. They are the lattice QCD, QCD sum rules, Schwinger--Dyson
approach, Nambu--Jona-Lasinio model, etc. In the present paper, we analyze the
photon-pion transition form factor in the gauged nonlocal chiral quark model
based on the picture of nontrivial QCD vacuum. The attractive feature of this
model is, that it interpolates the physics at large and small distances. At
low energy, it enjoys the spontaneous breaking of chiral symmetry, the
generation of the dynamical quark mass, and it satisfies the basic low energy
theorems. At energies much higher than the characteristic hadronic scale, it
becomes the theory of free massless quarks (in chiral limit).

The paper is organized as follows: In Sec. II, we give the basic elements of
the effective chiral quark model, the quark propagator and the quark-photon
and quark-pion vertices. In Sec. III, we transform the expression for the pion
transition form factor into the $\alpha$-representation and analyze, under
rather general requirements on the nonperturbative dynamics, the asymptotic
behavior of the form factor for different kinematics. Considering the
kinematics when one photon is virtual and other is real, we show that two
possible behaviors of the quark-pion vertex at large quark virtualities
results in two different asymptotic regimes for the pion form factor. One of
them corresponds to a standard factorized scheme with actual ${1/}Q^{2}$
asymptotics. The other provides a nonstandard asymptotic regime leading to
$\sim\ln{\left(  Q^{2}\right)  /}Q^{2}$ large-$Q^{2}$ behavior of the pion
form factor.
In Sec. IV, we specify two kinds of nonlocal chiral quark model implementing
different asymptotic regimes and obtain the pion DA for various sets of
parameters. In Sec. V, we are looking for the space of parameters that give a
satisfactory fit of the CELLO, CLEO and BABAR data. Sec. VI contains our conclusions.

\section{Nonlocal chiral quark model}

Let us discuss the properties of the triangle diagram (Fig. \ref{TriangleHard}%
) within the effective approach to nonperturbative QCD\ dynamics. To consider
the asymptotics of the photon-pion transition form factor, we do not need to
completely specify the elements of the diagram technique, which are, in
general, model dependent, but shall restrict ourselves to rather general
requirements. All expressions will be treated in Euclidean space appropriate
for the nonperturbative physics. The nonperturbative quark propagator, dressed
by the interaction with the QCD vacuum, is%
\begin{equation}
S\left(  k\right)  =\frac{\widehat{k}+m\left(  k^{2}\right)  }{D\left(
k^{2}\right)  }.\label{Qprop}%
\end{equation}
The main requirement to the quark propagator is, that at large quark
virtualities one has%
\begin{equation}
S\left(  k\right)  \overset{k^{2}\rightarrow\infty}{\rightarrow}\frac
{\widehat{k}}{k^{2}}.\label{QpropAS}%
\end{equation}
We assume also, that the dynamical quark mass is a function of the quark
virtuality $k^{2}$ and normalized at zero as%
\begin{equation}
m\left(  0\right)  =M_{q},\qquad D\left(  0\right)  =M_{q}^{2}.\label{M0}%
\end{equation}
At large virtualities, it drops to the current quark mass $m_{curr}$ faster
than any power of $k^{-2}$ (see the discussion in \cite{Dorokhov:2005pg})%
\begin{equation}
m\left(  k^{2}\right)  \sim M_{q}\exp\left(  -\left(  k^{2}\right)
^{a}\right)  +m_{curr},\quad a>0.\label{DynMassAs}%
\end{equation}
This is, firstly, because the dynamical quark mass is directly related to the
nonlocal quark condensate \cite{Dorokhov:1997iv,Dorokhov:2000gu} and,
secondly, the quark propagators with powerlike dynamical mass induce false
power corrections that are in contradiction to OPE. On the other hand, the
dynamical quark mass (\ref{DynMassAs}) generates exponentially small
corrections, invisible in the standard OPE. The direct instanton contributions
provide a famous example of these exponential corrections in the QCD sum rules
approach \cite{Shuryak:1982qx,Dorokhov:1989zw}. The denominator in
(\ref{Qprop}) at large virtualities is $D\left(  k^{2}\right)  \overset
{k^{2}\rightarrow\infty}{\rightarrow}k^{2}$ and the typical expression is
\begin{equation}
D\left(  k^{2}\right)  =k^{2}+m^{2}\left(  k\right)  .\label{QDen}%
\end{equation}

It is well known (see, e.g., \cite{Terning:1991yt,Bowler:1994ir}), that the
change of the quark propagator leads to a modification of the quark-photon
vertex in order to preserve the Ward-Takahashi identity
\begin{equation}
\Gamma_{\mu}\left(  k,q,k^{\prime}=k+q\right)  =-ie_{q}\left[  \gamma^{\mu
}-\Delta\Gamma_{\mu}\left(  k,q,k^{\prime}=k+q\right)  \right]
,\label{QGammaVert}%
\end{equation}
where the extra term guarantees the property%
\begin{equation}
q_{\mu}\Gamma_{\mu}\left(  k,q,k^{\prime}=k+q\right)  =S^{-1}\left(
k^{\prime}\right)  -S^{-1}\left(  k\right)  .\label{WTI}%
\end{equation}
The term $\Delta\Gamma_{\mu}\left(  q\right)  $ is not uniquely defined, even
within a particular model, especially its transverse part. The importance of
the full vertex $\Gamma_{\mu}$ is, that the axial anomaly is reproduced
\cite{Plant:2000ty}, and thus the photon-transition form factor correctly
normalized. Fortunately, due to the fact, that $\Delta\Gamma_{\mu}$ is not
proportional to $\gamma_{\mu}$ matrix, the corresponding amplitude has no
projection onto the leading twist operator. Thus, this term is suppressed, if
a large photon virtuality passes through the vertex, and hence does not
participate in the leading asymptotics of the form factor. Its leading
asymptotics results exclusively from the local part of the photon vertex%
\begin{equation}
\Gamma_{\mu}^{\mathrm{As}}\left(  k,q,k^{\prime}=k+q\right)  =-ie_{q}%
\gamma^{\mu}.\label{QGammaVertAs}%
\end{equation}

Furthermore, we need the quark-pion vertex,
\begin{equation}
\Gamma_{\pi}^{a}\left(  p\right)  =\frac{i}{f_{\pi}}\gamma_{5}\tau^{a}F\left(
k_{+}^{2},k_{-}^{2}\right)  ,\label{QPiVert}%
\end{equation}
where $k_{+}$ and $k_{-}$ are the quark and antiquark momenta. It is important
to note, that the quark-pion vertex function $F\left(  k_{+}^{2},k_{-}%
^{2}\right)  $ plays a similar role in our consideration as the light-cone
wave function $\Psi\left(  x,k_{\bot}\right)  $ in
\cite{Radyushkin:1977gp,Lepage:1979zb,Efremov:1978rn,Efremov:1979qk,Efremov:1978fi,Lepage:1980fj,Brodsky:1981rp}%
. The vertex function $F\left(  k_{+}^{2},k_{-}^{2}\right)  $ is symmetric in
the quark virtualities $k_{+}^{2}$ and $k_{-}^{2}$, and rapidly decreases,
when both virtualities are large. If it were a function of a linear
combination of the quark momenta $k_{+}$ and $k_{-}$, then it would led to a
growing form factor with increasing spacelike photon momenta (see for
discussions \cite{Dorokhov:2002iu}). The spontaneous breaking of chiral
symmetry ensures, that the vertex function $F\left(  k_{+}^{2},k_{-}%
^{2}\right)  $ is a functional of the dynamical mass $m\left(  k^{2}\right)
$. In particular, the vertex function is normalized via%
\begin{equation}
F\left(  k^{2},k^{2}\right)  =m\left(  k^{2}\right)  .\label{Fnorm}%
\end{equation}

In the following, the important feature of the vertex function $F\left(
k_{+}^{2},k_{-}^{2}\right)  $ will be its behavior in the limit, when one
quark virtuality is asymptotically large and the other remains finite. There
are two possibilities,%
\begin{equation}
F^{f}\left(  k_{+}^{2},k_{-}^{2}\right)  \overset{k_{-}^{2}\rightarrow\infty
}{\rightarrow}0, \label{QPiVertIAs}%
\end{equation}
and
\begin{equation}
F^{uf}\left(  k_{+}^{2},k_{-}^{2}\right)  \overset{k_{-}^{2}\rightarrow\infty
}{\rightarrow}g\left(  k_{+}^{2}\right)  . \label{QPiVertTAs}%
\end{equation}

Finally, one needs the projection of the pion state onto the leading twist
operator, see Fig. \ref{FactBL},
\begin{equation}
\Gamma_{\mu}^{5,\mathrm{As}}\left(  k,q,k^{\prime}=k+q\right)  =\gamma^{\mu
}\gamma^{5}.\label{QAVertAs}%
\end{equation}
This projection is determined by the matrix element $\left\langle 0\left\vert
\overline{q}\gamma^{\mu}\gamma^{5}\tau^{a}q\right\vert \pi^{a}\left(
p\right)  \right\rangle =-i2f_{\pi,\mathrm{PS}}$, where the constant
$f_{\pi,\mathrm{PS}}$ is (here $m^{\prime}\left(  u\right)  =dm\left(
u\right)  /du$)%
\begin{equation}
f_{\pi,\mathrm{PS}}^{2}=\frac{N_{c}}{4\pi^{2}}\int_{0}^{\infty}du\quad
u\frac{F\left(  u,u\right)  }{D^{2}\left(  u\right)  }\left(  m\left(
u\right)  -\frac{1}{2}um^{\prime}\left(  u\right)  \right)  ,\label{FpiPS}%
\end{equation}
which coincides with the square of the pion decay constant $f_{\pi
,\mathrm{PS}}$ in the so-called Pagels-Stokar form \cite{Pagels:1979hd}.
However note, that the physical pion decay constant, $f_{\pi},$ entering the
pion vertex (\ref{QPiVert}), is calculated by using the axial vertex
corresponding to the conserved axial current $\Gamma_{\mu}^{5}\left(
q\right)  $. It turns out that the constant $f_{\pi,\mathrm{PS}}$ and the
physical decay constant $f_{\pi}$ are not always identical. We return to this
point in Sec. IV.

Thus, we emphasize again, that in order to analyze the asymptotic behavior of
the pion transition form factor $F_{\pi\gamma^{\ast}\gamma^{\ast}}(Q_{1}%
^{2},Q_{2}^{2})$ by inspecting the triangular diagram, one needs to specify
only very general properties of the transition, from soft to hard regimes of
the quark-pion-photon dynamics encoded in (\ref{QpropAS}), (\ref{QGammaVertAs}%
), (\ref{QPiVert}) and (\ref{FpiPS}). At the same time, the full dynamics
(\ref{Qprop}), (\ref{QGammaVert}), (\ref{QPiVert}) should guarantee the low
energy theorems, in particular, the correct normalization of the form factor
by the axial anomaly
\begin{equation}
F_{\pi\gamma\gamma}(0,0)=1/\left(  4\pi^{2}f_{\pi}\right)  ,
\label{AxialAnomaly}%
\end{equation}
and the Goldberger-Treiman relation, connecting the quark-pion coupling
$g_{q\pi}$ and the dynamical quark mass $M_{q}$ with the physical pion decay
constant $f_{\pi}$: $f_{\pi}=M_{q}/g_{q\pi}$.

\section{Asymptotics of pion-photon transition form factor}

The invariant amplitude for the process $\gamma^{\ast}\gamma^{\ast}%
\rightarrow\pi^{0}$ is given by
\begin{equation}
A\left(  \gamma^{\ast}\left(  q_{1},\epsilon_{1}\right)  \gamma^{\ast}\left(
q_{2},\epsilon_{2}\right)  \rightarrow\pi^{0}\left(  p\right)  \right)
=-ie^{2}\varepsilon_{\mu\nu\rho\sigma}\epsilon_{1}^{\mu}\epsilon_{2}^{\nu
}q_{1}^{\rho}q_{2}^{\sigma}F_{\pi\gamma^{\ast}\gamma^{\ast}}\left(  -q_{1}%
^{2},-q_{2}^{2}\right)  ,\label{Ampl}%
\end{equation}
where $\epsilon_{i}^{\mu}$ are the photon polarization vectors, $p^{2}=m_{\pi
}^{2},q_{1}^{2}=-Q_{1}^{2},q_{2}^{2}=-Q_{2}^{2}$. In the effective nonlocal
quark-model considered above, one finds the contribution of the triangle
diagram to the invariant amplitude \cite{Dorokhov:2002iu},%
\[
A\left(  p^{2};q_{1}^{2},q_{2}^{2}\right)  =A^{\mathrm{loc}}\left(
p^{2};q_{1}^{2},q_{2}^{2}\right)  +A^{\mathrm{nonloc}}\left(  p^{2};q_{1}%
^{2},q_{2}^{2}\right)  ,
\]
where the first term contains only local part of the photon vertices
\begin{align}
&  A^{\mathrm{loc}}\left(  p^{2};q_{1}^{2},q_{2}^{2}\right)  =-ie^{2}%
\frac{N_{c}}{3f_{\pi}}\int\frac{d^{4}k}{(2\pi)^{4}}F(-k_{+}^{2},-k_{-}%
^{2})\label{MPiGG}\\
&  \cdot\left\{  tr[i\gamma_{5}S(k_{-})\widehat{\epsilon}_{2}S\left(
k_{+}-q_{1}\right)  ]\widehat{\epsilon}_{1}S(k_{+})]+\left(  q_{1}%
\leftrightarrow q_{2};\epsilon_{1}\leftrightarrow\epsilon_{2}\right)
\right\}  ,\nonumber
\end{align}
and the second term comprises the rest%
\begin{equation}
A^{\mathrm{nonloc}}\left(  p^{2};q_{1}^{2},q_{2}^{2}\right)  =-ie^{2}%
\frac{N_{c}}{3f_{\pi}}\int\frac{d^{4}k}{(2\pi)^{4}}F(-k_{+}^{2},-k_{-}%
^{2})\label{Mnonloc}%
\end{equation}%
\[
\cdot\left\{  tr[i\gamma_{5}S(k_{-})S\left(  k_{+}-q_{1}\right)
\widehat{\epsilon}_{1}S(k_{+})]\left(  \epsilon_{2},\Delta\Gamma\left(
k_{+},-q_{1},k_{+}-q_{1}\right)  \right)  \right.
\]%
\[
\left.  +tr[i\gamma_{5}S(k_{-})\widehat{\epsilon}_{2}S\left(  k_{+}%
-q_{1}\right)  S(k_{+})]\left(  \epsilon_{1},\Delta\Gamma\left(  k_{+}%
-q_{1},-q_{2},k_{-}\right)  \right)  \right\}  +\left(  q_{1}\leftrightarrow
q_{2};\epsilon_{1}\leftrightarrow\epsilon_{2}\right)  ,
\]
with $p=q_{1}+q_{2},$ $q=q_{1}-q_{2},$ $k_{\pm}=k\pm p/2$.

As we discussed above, the leading asymptotics results from the local part of
the amplitude, $A^{\mathrm{loc}}$. After taking the Dirac trace and going to
Euclidian metric $\left(  d^{4}k\rightarrow id^{4}k,k^{2}\rightarrow
-k^{2}\right)  $, one obtains%
\begin{equation}
A^{\mathrm{loc}}\left(  p^{2};q_{1}^{2},q_{2}^{2}\right)  =\frac{e^{2}N_{c}%
}{6\pi^{2}f_{\pi}}\int\frac{d^{4}k}{\pi^{2}}F(k_{+}^{2},k_{-}^{2}%
)\frac{m\left(  k_{+}^{2}\right)  \left(  \varepsilon_{12kq_{2}}%
-\varepsilon_{12q_{1}q_{2}}\right)  -m\left(  k_{-}^{2}\right)  \varepsilon
_{12q_{1}k}+m\left(  k_{3}^{2}\right)  \varepsilon_{12pk}}{D\left(  k_{+}%
^{2}\right)  D\left(  k_{-}^{2}\right)  D\left(  k_{3}^{2}\right)  },
\label{AlocK}%
\end{equation}
where $k_{3}^{2}=\left(  k_{+}-q_{1}\right)  ^{2}$, and $\varepsilon
_{12kq_{2}}=\varepsilon_{\mu\nu\lambda\rho}\epsilon_{1}^{\mu}\epsilon_{2}%
^{\nu}k^{\lambda}q_{2}^{\rho}$, etc.

In order to analyze the asymptotic properties of the form factor, let us
transform the integral in (\ref{AlocK}) formally into the $\alpha$
representation (see \cite{Bogolyubov:1980,Zavialov:1990}), which is one of the
basic methods for the study of hard processes in perturbative QCD
\cite{Radyushkin:1997ki}, as well as in nonperturbative quark models
\cite{Dorokhov:2000gu}. Let us define for any function $F$ of virtuality
$k^{2}$, decaying at large virtuality as $1/k^{2}$ or faster, its $\alpha$
representation (Laplace transform)
\begin{equation}
F\left(  k^{2}\right)  =\int_{0}^{\infty}d\alpha e^{-\alpha k^{2}}f\left(
\alpha\right)  ,\qquad F\left(  k^{2}\right)  \sim f\left(  \alpha\right)  ,
\label{AlphaDef}%
\end{equation}
where $F\left(  k^{2}\right)  $ is the image of the original $f\left(
\alpha\right)  $. The important asymptotic property of this representation is,
that the large power-like $k^{2}$ behavior of $F\left(  k^{2}\right)  $ is
given by derivatives of the original $g\left(  \alpha\right)  $ at $\alpha=0$%
\begin{equation}
F\left(  k^{2}\right)  \overset{k^{2}\rightarrow\infty}{=}\frac{f\left(
0\right)  }{k^{2}}+\frac{f^{\prime}\left(  0\right)  }{k^{4}}+\frac
{f^{\prime\prime}\left(  0\right)  }{k^{6}}+... \label{AlphaDefAs}%
\end{equation}
Thus, the large $k^{2}$ asymptotics of the image $F\left(  k^{2}\right)  $ is
related to the small $\alpha$ behavior of the original $f\left(
\alpha\right)  .$

Let us introduce the following notations
\begin{align}
&  \frac{1}{D\left(  k^{2}\right)  }\sim d\left(  \alpha\right)  ,\qquad
\frac{m\left(  k^{2}\right)  }{D\left(  k^{2}\right)  }\sim d_{m}\left(
\alpha\right)  ,\label{Dalpha}\\
&  \frac{F(k_{+}^{2},k_{-}^{2})}{D\left(  k_{+}^{2}\right)  D\left(  k_{-}%
^{2}\right)  }\sim G\left(  \alpha,\beta\right)  ,\quad\frac{m\left(
k_{+}^{2}\right)  F(k_{+}^{2},k_{-}^{2})}{D\left(  k_{+}^{2}\right)  D\left(
k_{-}^{2}\right)  }\sim G_{m,0}\left(  \alpha,\beta\right)  ,\label{Falpha}%
\end{align}
where in the second line the double $\alpha$ representation is implied.
Because of the properties (\ref{QpropAS}) and (\ref{DynMassAs}) one has%
\begin{align*}
d\left(  0\right)   &  =1,d^{\prime}\left(  0\right)  =0,d^{\prime\prime
}\left(  0\right)  =0,...,\\
d_{m}\left(  0\right)   &  =0,d_{m}^{\prime}\left(  0\right)  =0,...
\end{align*}

With this notation, using the standard technique of the $\alpha$
representation (\cite{Bogolyubov:1980,Zavialov:1990}), the momentum integral
in (\ref{AlocK}) is transformed into the following expression for the form
factor%
\begin{align}
F_{\pi\gamma^{\ast}\gamma^{\ast}}^{\mathrm{loc}}\left(  p^{2};Q_{1}^{2}%
,Q_{2}^{2}\right)   &  =\frac{N_{c}}{6\pi^{2}f_{\pi}}\int\frac{d\left(
\alpha\beta\gamma\right)  }{\Delta^{3}}e^{-\frac{1}{\Delta}\left[
-\alpha\beta p^{2}+\gamma\left(  \alpha Q_{1}^{2}+\beta Q_{2}^{2}\right)
\right]  }\label{AlocA}\\
&  \cdot\left[  d\left(  \gamma\right)  \left(  \alpha G_{m,0}\left(
\alpha,\beta\right)  +\beta G_{0,m}\left(  \alpha,\beta\right)  \right)
+\gamma d_{m}\left(  \gamma\right)  G\left(  \alpha,\beta\right)  \right]
,\nonumber
\end{align}
where $\Delta=\alpha+\beta+\gamma$ and $\int d\left(  \alpha\beta
\gamma\right)  ...=\int_{0}^{\infty}d\alpha\int_{0}^{\infty}d\beta\int
_{0}^{\infty}d\gamma...$

\subsection{Symmetric kinematics}

Let us first consider the symmetric kinematics $Q_{1}^{2}=Q_{2}^{2}=Q^{2}.$
Then one has%
\begin{align}
F_{\pi\gamma^{\ast}\gamma^{\ast}}^{\mathrm{loc}}\left(  p^{2};Q^{2}%
,Q^{2}\right)   &  =\frac{N_{c}}{6\pi^{2}f_{\pi}}\int\frac{d\left(
\alpha\beta\gamma\right)  }{\Delta^{3}}e^{-\frac{1}{\Delta}\left[
-\alpha\beta p^{2}+\gamma\left(  \alpha+\beta\right)  Q^{2}\right]
}\label{FpiSym}\\
&  \cdot\left[  d\left(  \gamma\right)  \left(  \alpha G_{m,0}\left(
\alpha,\beta\right)  +\beta G_{0,m}\left(  \alpha,\beta\right)  \right)
+\gamma d_{m}\left(  \gamma\right)  G\left(  \alpha,\beta\right)  \right]
.\nonumber
\end{align}
Large $Q^{2}$ behavior of $F_{\pi\gamma^{\ast}\gamma^{\ast}}^{\mathrm{loc}%
}\left(  p^{2};Q^{2},Q^{2}\right)  $ corresponds to either small $\gamma,$
small $\alpha+\beta$, or to large $\Delta.$ It is easy to check, that the
leading asymptotics is ensured by small $\gamma$ and thus $\Delta
\rightarrow\alpha+\beta$. The term with factor $\gamma d_{m}\left(
\gamma\right)  $ provides only exponentially small corrections and does not
contribute to the leading asymptotics. In this way, in (\ref{FpiSym}), the
integral over $\gamma$ (small distances) and the integral over $\alpha,\beta$
(large distances) is factorized. The integral over $\gamma$, using
(\ref{AlphaDef}), transforms the original $d\left(  \gamma\right)  $ back to
momentum space $1/D\left(  Q^{2}\right)  $%
\begin{equation}
F_{\pi\gamma^{\ast}\gamma^{\ast}}^{\mathrm{loc}}\left(  p^{2};Q^{2}%
,Q^{2}\right)  \overset{Q^{2}\rightarrow\infty}{=}\frac{N_{c}}{6\pi^{2}f_{\pi
}}\frac{1}{D\left(  Q^{2}\right)  }\int\frac{d\left(  \alpha\beta\right)
}{\left(  \alpha+\beta\right)  ^{3}}e^{\frac{\alpha\beta}{\alpha+\beta}p^{2}%
}\left(  \alpha G_{m,0}\left(  \alpha,\beta\right)  +\beta G_{0,m}\left(
\alpha,\beta\right)  \right)  .\label{FpiSymFact}%
\end{equation}
For the quark propagator, one obtains in this limit $1/D\left(  Q^{2}\right)
\rightarrow1/Q^{2}$ plus exponentially small corrections, due to the
properties (\ref{QpropAS}) and (\ref{DynMassAs}). It turns out, that the
integral in (\ref{FpiSymFact}) is the $\alpha$ representation of the pion
decay constant (\ref{FpiPS})%
\begin{equation}
f_{\mathrm{PS},\pi}^{2}=\frac{N_{c}}{4\pi^{2}}\int\frac{d\left(  \alpha
\beta\right)  }{\left(  \alpha+\beta\right)  ^{3}}e^{\frac{\alpha\beta}%
{\alpha+\beta}p^{2}}\left(  \alpha G_{m,0}\left(  \alpha,\beta\right)  +\beta
G_{0,m}\left(  \alpha,\beta\right)  \right)  .\label{FpiPSalpha}%
\end{equation}
Thus one obtains the asymptotic formula
\begin{equation}
F_{\pi\gamma^{\ast}\gamma^{\ast}}^{\mathrm{loc}}\left(  0;Q^{2},Q^{2}\right)
\overset{Q^{2}\rightarrow\infty}{=}\frac{2}{3}\frac{1}{Q^{2}}\frac
{f_{\mathrm{PS},\pi}^{2}}{f_{\pi}},\label{FpiSymFactChiral}%
\end{equation}
for the form factor in symmetric kinematics, which for models, where
$f_{\mathrm{PS},\pi}=f_{\pi}$, reproduces the Brodsky-Lepage factorization
result \cite{Brodsky:1981rp}. \begin{figure}[h]
\hspace*{-10mm}\includegraphics[width=0.6\textwidth]{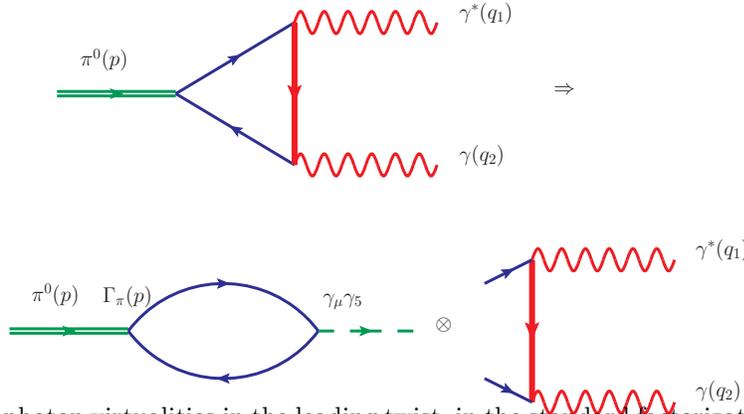}
\vspace*{-10mm}\caption{{\protect\footnotesize At large photon virtualities in
the leading twist, in the standard factorization regime, the amplitude is
factorized into the soft pion matrix element and the hard coefficient
function. }}%
\label{FactBL}%
\end{figure}

In order to define the pion DA, we carry out a change of variables in
(\ref{FpiPSalpha})%
\begin{equation}
\alpha\rightarrow xL,\qquad\beta\rightarrow\overline{x}L,\label{ABtoXL}%
\end{equation}
with $\overline{x}=\left(  1-x\right)  ,$ then%
\begin{equation}
\varphi_{\pi}\left(  x\right)  =\frac{N_{c}}{4\pi^{2}f_{\mathrm{PS},\pi}^{2}%
}\int_{0}^{\infty}\frac{dL}{L}e^{x\overline{x}Lp^{2}}\left(  xG_{m,0}\left(
xL,\overline{x}L\right)  +\overline{x}G_{0,m}\left(  xL,\overline{x}L\right)
\right)  ,\label{PionDAsym}%
\end{equation}
with%
\[
\int_{0}^{1}dx\varphi_{\pi}\left(  x\right)  =1.
\]
In the momentum representation and using the chiral limit $p^{2}=0$, the
result (\ref{PionDAsym}) for the leading twist DA is \cite{Dorokhov:2002iu}
\begin{equation}
\varphi_{\pi}(x)=\frac{N_{c}}{4\pi^{2}f_{\mathrm{PS},\pi}^{2}}\int_{-\infty
}^{\infty}\frac{d\lambda}{2\pi}\int_{0}^{\infty}du\frac{F(u+i\lambda
\overline{x},u-i\lambda x)}{D\left(  u-i\lambda x\right)  D\left(
u+i\lambda\overline{x}\right)  }\left[  xm\left(  u+i\lambda\overline
{x}\right)  +\overline{x}m\left(  u-i\lambda x\right)  \right]  .
\end{equation}
\ This result is also in agreement with earlier calculations made in the
instanton model under some simplified assumptions
\cite{Esaibegian:1989uj,Dorokhov:1991nj,Petrov:1998kg,Anikin:1999cx}. The
arguments in the integrand have the simple meaning of the transverse $u\equiv
k_{\perp}^{2}$ and longitudinal parts of the quark (antiquark) virtualities.

For the pion vertex with the property (\ref{QPiVertIAs}), the pion DA vanishes
at the endpoints%
\[
\varphi_{\pi}^{f}\left(  x=0\right)  =\varphi_{\pi}^{f}\left(  x=1\right)  =0,
\]
while for the second type of the pion vertex (\ref{QPiVertTAs}), one has
instead%
\begin{equation}
\varphi_{\pi}^{uf}\left(  x=0\right)  =\varphi_{\pi}^{uf}\left(  x=1\right)
=\frac{N_{c}}{4\pi^{2}f_{\mathrm{PS},\pi}^{2}}\int_{0}^{\infty}du\frac
{m\left(  u\right)  g\left(  u\right)  }{D\left(  u\right)  }.\label{DAunf0}%
\end{equation}

The pion DA $\varphi_{\pi}\left(  x\right)  $ in (\ref{PionDAsym}) is the
leading twist-2 DA, defined as a gauge-invariant matrix element of the
nonlocal operator
\begin{equation}
\left\langle 0\left\vert \overline{d}\left(  z\right)  \gamma_{\mu}\gamma
_{5}P\exp\left(  \int_{-z}^{z}dz^{\mu}A_{\mu}\left(  z\right)  \right)
u\left(  -z\right)  \right\vert \pi^{+}\left(  p\right)  \right\rangle
=i\sqrt{2}f_{\pi}^{\mathrm{PS}}p_{\mu}\int_{0}^{1}dxe^{i\left(  2x-1\right)
pz}\varphi_{\pi}(x), \label{PiDAdefQCD}%
\end{equation}
with the Dirac structure $\gamma_{\mu}\gamma_{5}$ between the pion and vacuum
states, $z$ a light-like four-vector ($z^{2}=0$), and the gluon field $A_{\mu
}\left(  z\right)  $.

Thus, in symmetric kinematics, the standard factorization is not violated and
the OPE is modified only by exponentially small terms.

\subsection{Asymmetric kinematics I}

Let us now consider the asymmetric kinematics $Q_{1}^{2}=Q^{2},Q_{2}^{2}=0.$
Then one has%
\begin{align}
F_{\pi\gamma^{\ast}\gamma}^{\mathrm{loc}}\left(  p^{2};Q^{2},0\right)   &
=\frac{N_{c}}{6\pi^{2}f_{\pi}}\int\frac{d\left(  \alpha\beta\gamma\right)
}{\Delta^{3}}e^{-\frac{1}{\Delta}\left[  -\alpha\beta p^{2}+\gamma\alpha
Q^{2}\right]  }\label{FpiAsym}\\
&  \cdot\left[  d\left(  \gamma\right)  \left(  \alpha G_{m,0}\left(
\alpha,\beta\right)  +\beta G_{0,m}\left(  \alpha,\beta\right)  \right)
+\gamma d_{m}\left(  \gamma\right)  G\left(  \alpha,\beta\right)  \right]
.\nonumber
\end{align}
For simplicity in the following we shell consider the chiral limit,
$m_{curr}=0,p^{2}=0.$

Let us first consider the model with the quark-pion vertex possessing the
property (\ref{QPiVertIAs}). In this case, the regime of small $\alpha$ does
not lead to the leading asymptotic terms because of property $G\left(
\alpha,\beta\right)  \rightarrow0$ as $\alpha\rightarrow0$. The leading large
$Q^{2}$ behavior corresponds to small $\gamma,$ i.e. $\Delta\rightarrow
\alpha+\beta$, (Fig. \ref{TriangleHard}a), as for symmetric kinematics,
\begin{align}
&  F_{\pi\gamma^{\ast}\gamma}^{\mathrm{loc,I}}\left(  0;Q^{2},0\right)
\overset{Q^{2}\rightarrow\infty}{=}\label{FpiAsymAs}\\
&  \frac{N_{c}}{6\pi^{2}f_{\pi}}\int\frac{d\left(  \alpha\beta\gamma\right)
}{\left(  \alpha+\beta\right)  ^{3}}e^{-\frac{1}{\alpha+\beta}\gamma\alpha
Q^{2}}d\left(  \gamma\right)  \left(  \alpha G_{m,0}\left(  \alpha
,\beta\right)  +\beta G_{0,m}\left(  \alpha,\beta\right)  \right)  .\nonumber
\end{align}
This asymptotic term corresponds to the standard factorization contribution
(Fig. \ref{FactBL}) and the integral over $\gamma$ again can be transformed
back to the momentum space
\[
F_{\pi\gamma^{\ast}\gamma}^{\mathrm{loc,I}}\left(  0;Q^{2},0\right)
\overset{Q^{2}\rightarrow\infty}{=}\frac{N_{c}}{6\pi^{2}f_{\pi}}\int
\frac{d\left(  \alpha\beta\right)  }{\left(  \alpha+\beta\right)  ^{3}}%
\frac{\alpha G_{m,0}\left(  \alpha,\beta\right)  +\beta G_{0,m}\left(
\alpha,\beta\right)  }{D\left(  \frac{\alpha Q^{2}}{\alpha+\beta}\right)  }.
\]
After change of variables (\ref{ABtoXL}), we arrive at the representation%
\begin{equation}
F_{\pi\gamma^{\ast}\gamma}^{\mathrm{loc,I}}\left(  0;Q^{2},0\right)
\overset{Q^{2}\rightarrow\infty}{=}\frac{2}{3}\frac{f_{\mathrm{PS},\pi}^{2}%
}{f_{\pi}}\int_{0}^{1}dx\frac{1}{D\left(  xQ^{2}\right)  }\varphi_{\pi}%
^{f}\left(  x\right)  ,\label{FpqcdNL}%
\end{equation}
where $\varphi_{\pi}\left(  x\right)  $ is defined in (\ref{PionDAsym}).
Because in the considered case $\varphi_{\pi}\left(  x\right)  $ vanishes at
the endpoints the actual asymptotics is%
\begin{equation}
F_{\pi\gamma^{\ast}\gamma}^{As\mathrm{,I}}\left(  0;Q^{2},0\right)
\overset{Q^{2}\rightarrow\infty}{=}\frac{1}{Q^{2}}\frac{2}{3}\frac
{f_{\mathrm{PS},\pi}^{2}}{f_{\pi}}J^{f}\label{FNLac}%
\end{equation}
in agreement with (\ref{Fpqcd}), where $J^{f}=\int_{0}^{1}\frac{dx}{x}%
\varphi_{\pi}^{f}\left(  x\right)  $ is given in the momentum space
representation as \cite{Dorokhov:2002iu}%
\begin{equation}
J^{f}=\frac{N_{c}}{4\pi^{2}f_{\mathrm{PS},\pi}^{2}}\int_{0}^{\infty}du\frac
{u}{D\left(  u\right)  }\int_{0}^{1}dy\frac{F^{f}\left(  u,yu\right)  m\left(
yu\right)  }{D\left(  yu\right)  }.\label{InvMom}%
\end{equation}

As we have already noted in Introduction the asymptotic behavior (\ref{FNLac})
is not seen in the BABAR data. Nevertheless, even for the case considered, in
principle, it is possible to simulate in some wide preasymptotic kinematical
region a logarithmically enhanced behavior of the form factor. This happens if
one assumes that the pion DA entering (\ref{FpqcdNL}) is almost flat
$\varphi_{\pi}\left(  x\right)  \approx1,$\ i.e. it is close to a constant
everywhere except small vicinity near endpoints. Then, in order to regularize
the integral for $J^{f}$ in the infrared region, one needs to keep the
exponentially small terms in (\ref{FpqcdNL}).

To this end, let us analyze the asymptotic behavior of the integral
\begin{equation}
J^{L}=Q^{2}\int_{0}^{1}dx\frac{1}{D\left(  xQ^{2}\right)  },\label{JL}%
\end{equation}
corresponding to a flat pion DA, for some popular models of the
nonperturbative quark propagator. Firstly, we consider the quark propagator
\begin{equation}
\frac{1}{D\left(  k^{2}\right)  }=\frac{1-\exp\left(  -k^{2}/\Lambda
^{2}\right)  }{k^{2}}\label{QpropConf}%
\end{equation}
with the property of analytical confinement \cite{Efimov:1988yd,Efimov:1993zg}%
. In quark models, where this propagator is used, the parameter $\Lambda$ has
the meaning of a dynamical quark mass \cite{Radzhabov:2003hy}, $\Lambda\equiv
M_{q},$ with typical values of $M_{q}=200-300$ MeV. Inserting (\ref{QpropConf}%
) into (\ref{JL}) one obtains%
\begin{equation}
J_{AC}^{L}=\int_{0}^{1}dx\frac{1-\exp\left(  -xQ^{2}/M_{q}^{2}\right)  }%
{x}\label{FpqcdNLrad}%
\end{equation}
with the leading asymptotic behavior%
\begin{equation}
J_{AC}^{L}\overset{Q^{2}\rightarrow\infty}{=}\ln\left(  Q^{2}/M_{q}%
^{2}\right)  +\gamma_{E},\label{FpqcdNLAs}%
\end{equation}
where $\gamma_{E}$ is the Euler-Mascheroni constant. Both expressions
(\ref{FpqcdNLrad}) and (\ref{FpqcdNLAs}) are very close to the result
(\ref{Frad}) obtained in \cite{Radyushkin:2009zg}. The difference is, that in
the expression (\ref{FpqcdNLrad})  the extra factor $\left(  1-x\right)
^{-1}$ in the exponent is absent, and more important the parameter in the
exponent in  (\ref{FpqcdNLrad}) has clear physical sense as a dynamical quark
mass squared.

Secondly, let us take the propagator of the general form given in
(\ref{Qprop})%
\begin{equation}
J_{Q}^{L}=Q^{2}\int_{0}^{1}dx\frac{1}{xQ^{2}+m^{2}\left(  xQ^{2}\right)
}.\label{FpqcdNLdyn}%
\end{equation}
Then one obtains the asymptotic behavior
\begin{equation}
J_{Q}^{L}\overset{Q^{2}\rightarrow\infty}{=}\ln\left( Q^{2}/M_{q}%
^{2}\right)  +\int_{0}^{\infty}du\frac{M_{q}^{2}-m^{2}\left(  u\right)
}{\left(  u+m^{2}\left(  u\right)  \right)  \left(  u+M_{q}^{2}\right)
}.\label{JQas}%
\end{equation}
Again, this is similar to (\ref{FradM}) obtained in \cite{Radyushkin:2009zg},
but with important differences. In fact, (\ref{FradM}) is a purely asymptotic
formula and it is not allowed to keep the parameter $M$ in the asymptotic
quark propagator. The expression (\ref{FpqcdNLdyn}) is valid for all $Q^{2}$
and provides the leading asymptotics for the flat DA (\ref{JQas}). It has
correct large $Q^{2}$ behavior for the quark propagator, $1/Q^{2}$, and does
not contain false power corrections.

\subsection{Asymmetric kinematics II}

Now, let us consider the model with the quark-pion vertex possessing the
property (\ref{QPiVertTAs}). It is convenient to rearrange the terms in the
pion form factor in the following way%
\begin{align}
&  F_{\pi\gamma^{\ast}\gamma}^{\mathrm{loc,II}}\left(  0;Q^{2},0\right)
=\frac{N_{c}}{6\pi^{2}f_{\pi}}\int\frac{d\left(  \alpha\beta\gamma\right)
}{\Delta^{3}}e^{-\frac{\gamma\alpha}{\Delta}Q^{2}}\left\{  \beta r_{m}\left(
\beta\right)  \right.  \label{FlocII}\\
&  +\alpha G_{m,0}\left(  \alpha,\beta\right)  d\left(  \gamma\right)
+\beta\left[  G_{0,m}\left(  \alpha,\beta\right)  -r_{m}\left(  \beta\right)
\right]  \nonumber\\
&  +\gamma G\left(  \alpha,\beta\right)  d_{m}\left(  \gamma\right)  +\beta
r_{m}\left(  \beta\right)  \left[  d\left(  \gamma\right)  -1\right]
\nonumber\\
&  \left.  +\beta\left[  d\left(  \gamma\right)  -1\right]  \left[  \beta
G_{0,m}\left(  \alpha,\beta\right)  -r_{m}\left(  \beta\right)  \right]
\right\}  ,\nonumber
\end{align}
where we introduce notations for the originals%
\[
\frac{g\left(  k^{2}\right)  }{D\left(  k^{2}\right)  }\sim r\left(
\alpha\right)  ,\qquad\frac{m\left(  k^{2}\right)  g\left(  k^{2}\right)
}{D\left(  k^{2}\right)  }\sim r_{m}\left(  \alpha\right)  .
\]

The term in the fourth line of (\ref{FlocII}) vanishes as $\gamma\rightarrow0$
or $\alpha\rightarrow0$, and thus does not participate in the leading
asymptotics. The terms in the second line vanish as  $\alpha\rightarrow0$, but
remains finite as $\gamma\rightarrow0.$ In this case, the $1/Q^{2}$
asymptotics is due to hard quark propagator connecting two photon vertices and
the coefficient reflects the soft properties of the pion (Fig.
\ref{TriangleHard}a and Fig. 3). For the terms in the third line one has an
opposite situation, they vanish with $\gamma$, but finite as $\alpha
\rightarrow0$. Thus, the $1/Q^{2}$ asymptotics is due to hard quark propagator
connecting pion and hard photon vertices, while the coefficient correlates
soft properties of the pion and photon (Fig. \ref{TriangleHard}b and Fig. 4).
The term in the first line of (\ref{FlocII})  provides the asymptotics $\sim$
$\ln\left(  Q^{2}\right)  /Q^{2}.$  This asymptotics corresponds to a combined
soft-hard regime when one parameter (i.e. $\alpha$) vanishes, while the other
($\gamma$) goes to infinity\footnote{See for classification of different
regimes \cite{Radyushkin:1997ki}.}.

After standard manipulations with the integrals one obtains the following
large-$Q^{2}$ asymptotic behavior transformed to the momentum representation%
\begin{align}
& F_{\pi\gamma^{\ast}\gamma}^{As\mathrm{,II}}\left(  0;Q^{2},0\right)
\overset{Q^{2}\rightarrow\infty}{=}\frac{1}{Q^{2}}\frac{N_{c}}{6\pi^{2}f_{\pi
}}\left[  \int_{0}^{\infty}du\frac{m\left(  u\right)  g\left(  u\right)
}{D\left(  u\right)  }\ln\left(  \frac{Q^{2}}{u}\right)  +A\right]
,\label{FIIas}\\
& A=\int_{0}^{\infty}du\frac{1}{D\left(  u\right)  }\int_{0}^{1}%
dy\frac{m\left(  yu\right)  }{D\left(  yu\right)  }\left\{  uF^{uf}\left(
u,yu\right)  -\left[  u+2m^{2}\left(  u\right)  \right]  g\left(  yu\right)
\right\}  .\label{FIIasC}%
\end{align}
The coefficient of the logarithmic term in (\ref{FIIas}) is clearly related to
the fact that the pion DA for the case considered does not vanish at the
endpoints and proportional to the value of the pion DA at these points, see
(\ref{DAunf0}). When the function $g\left(  u\right)  \equiv0,$ we reproduce
the asymptotics (\ref{FNLac}) and (\ref{InvMom}) corresponding to the quark
pion vertex with property (\ref{QPiVertIAs}). The variable $u$ in the integral
(\ref{FIIas}) may be considered as the square of quark transverse momentum in
the pion, $u\sim k_{\perp}^{2}$. The asymptotic expression (\ref{FIIas})
generalizes the asymptotic formula (\ref{Fpqcd}) for the case when the
standard factorization is violated.

\begin{figure}[h]
\hspace*{-10mm}\includegraphics[width=0.6\textwidth]{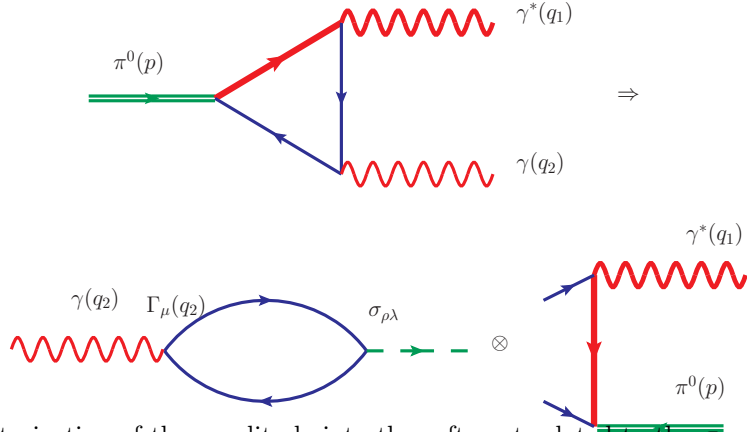}
\vspace*{-10mm}\caption{{\protect\footnotesize The factorization of the
amplitude into the soft part related to the }$\sigma_{\mu\nu}$%
{\protect\footnotesize \ projection of the photon wave function and the hard
part of the quark propagator.}}%
\end{figure}

\section{The instanton and chiral models}

In the previous section we considered the asymptotic behavior of the pion
transition form factor given in (\ref{Ampl})-(\ref{Mnonloc}). In order to
calculate this form factor in the whole kinematic region and compare with
available experimental data, we should further specify our model assumptions.
Let us introduce the momentum-dependent dynamical quark mass entering the
propagator (\ref{Qprop}) as $\left(  \text{we consider the chiral limit
}m_{curr}=0\right)  $%
\begin{equation}
m\left(  k^{2}\right)  =M_{q}f^{2}\left(  k^{2}\right)  \label{DynMass}%
\end{equation}
and take the profile function $f\left(  k^{2}\right)  $ in a Gaussian form%
\begin{equation}
f\left(  k^{2}\right)  =\exp\left(  -\Lambda k^{2}\right)  .\label{GaussProf}%
\end{equation}
Thus, the model contains two parameters, the dynamical quark mass $M_{q}$ and
the non-locality parameter $\Lambda.$

Next, we need to specify the nonlocal part of the vector vertex that does not
participate in the leading asymptotics, but is very important in implementing
the low energy theorems. The nonlocal part of the vector vertex in
(\ref{QGammaVert}) is taken of the form \cite{Terning:1991yt}%
\begin{equation}
\Delta\Gamma_{\mu}\left(  k,q,k^{\prime}=k+q\right)  =\left(  k+k^{\prime
}\right)  _{\mu}\frac{m\left(  k^{\prime2}\right)  -m\left(  k^{2}\right)
}{k^{\prime2}-k^{2}}. \label{DQGammaVert}%
\end{equation}

Further, we will consider two kinds of quark-pion vertex (\ref{QPiVert}), the
first given by
\begin{equation}
F_{I}\left(  k_{+}^{2},k_{-}^{2}\right)  =M_{q}f\left(  k_{+}^{2}\right)
f\left(  k_{-}^{2}\right)  , \label{QPiVertI}%
\end{equation}
and the second by%
\begin{equation}
F_{\chi}\left(  k_{+}^{2},k_{-}^{2}\right)  =\frac{1}{2}M_{q}\left[
f^{2}\left(  k_{+}^{2}\right)  +f^{2}\left(  k_{-}^{2}\right)  \right]  .
\label{QPiVertT}%
\end{equation}
The first one is motivated by the instanton picture of QCD vacuum
\cite{Diakonov:1985eg} and the second by the nonlocal chiral quark model
advertised in \cite{Holdom:1990iq}. We shall in the further discussion refer
to vertex function (\ref{QPiVertI}), which has the $k^{2}\rightarrow\infty$
behavior (\ref{QPiVertIAs}), as the instanton model, and to the other choice
(\ref{QPiVertT}), corresponding to $k^{2}\rightarrow\infty$
behavior(\ref{QPiVertTAs}), as the chiral model.

The important requirement, that correlates the parameters of the models, is to
fit the pion decay constant $f_{\pi}$. \ For the instanton based model this
constant is given by the expression found in \cite{Diakonov:1985eg}
\begin{equation}
f_{\mathrm{DP},\pi}^{2}=\frac{N_{c}}{4\pi^{2}}\int_{0}^{\infty}du\quad
u\frac{m\left(  u\right)  }{D^{2}\left(  u\right)  }\left(  m\left(  u\right)
-um^{\prime}\left(  u\right)  +u^{2}\left(  m^{\prime}\left(  u\right)
\right)  ^{2}\right)  , \label{FpiDP}%
\end{equation}
and for the chiral model \cite{Holdom:1990iq} the expression for $f_{\pi}$
coincides with the Pagels-Stokar form (\ref{FpiPS}). Within the nonlocal
chiral model approach there is a difference between the vertex corresponding
to the conserved axial current,
\begin{equation}
\Gamma_{\mu}^{5}\left(  k,q,k^{\prime}=k+q\right)  =\left[  \gamma^{\mu}%
\gamma^{5}-\Delta\Gamma_{\mu}^{5}\left(  k,q,k^{\prime}=k+q\right)  \right]
\label{QAVert}%
\end{equation}
and the local vertex (\ref{QAVertAs}), corresponding to the leading twist
operator. The total axial vertex $\Gamma_{\mu}^{5}\left(  q\right)  $ ensures
the axial Ward-Takahashi identity and the Goldberger-Treiman relation. The
nonlocal part of the axial vertex, that leads to (\ref{FpiPS}) is given in
\cite{Holdom:1992fn} and to (\ref{FpiDP}) is given in
\cite{Bowler:1994ir,Anikin:2000rq,Dorokhov:2003kf}.

\begin{figure}[th]
\hspace*{-10mm}\includegraphics[width=0.8\textwidth]{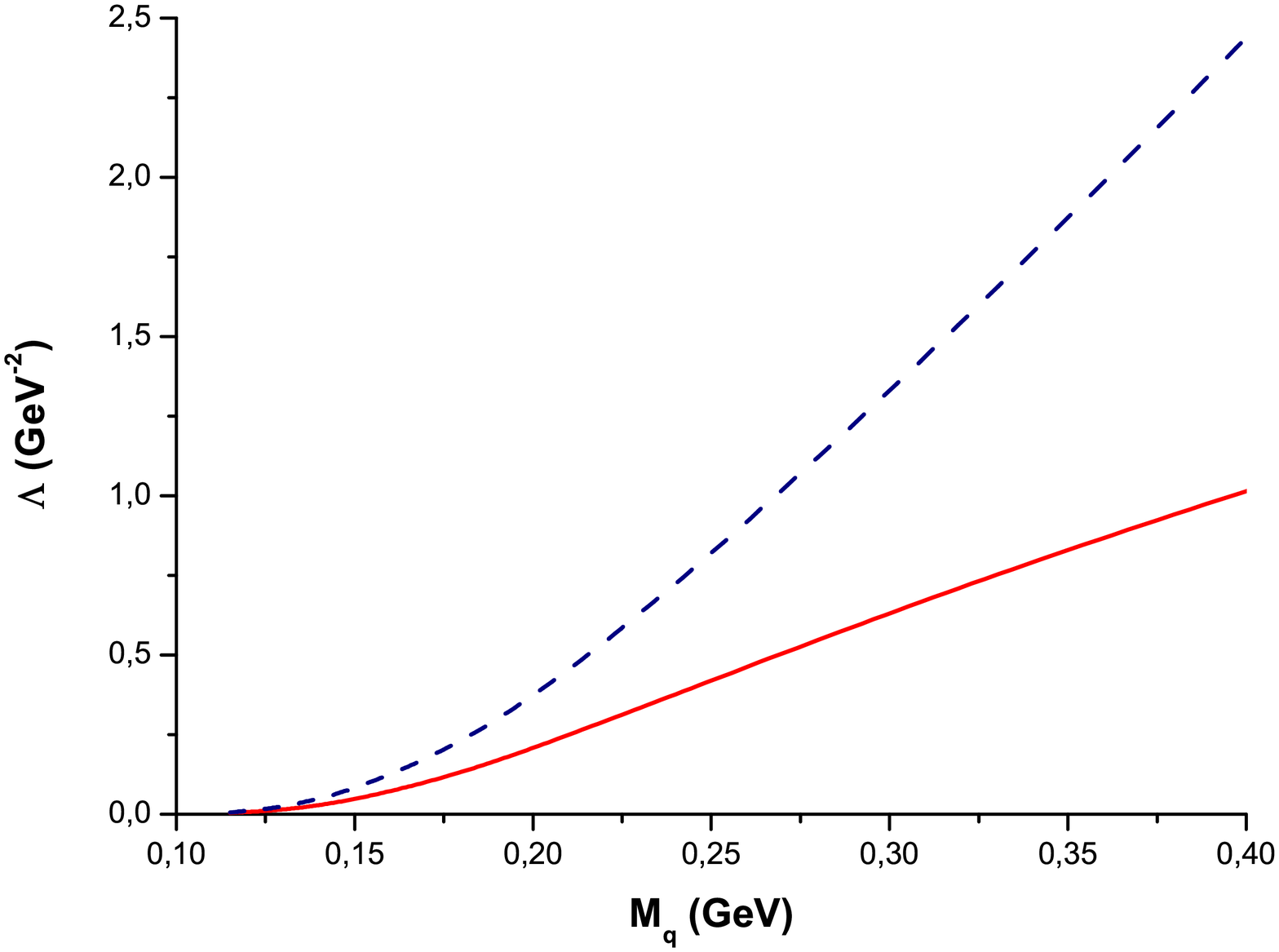}
\vspace*{-10mm}\caption{{\protect\footnotesize The correlation between
dynamical quark mass }$M_{q}${\protect\footnotesize \ and the nonlocal
parameter }$\Lambda${\protect\footnotesize \ that fit the pion decay constant
in chiral limit }$f_{\pi}=85${\protect\footnotesize \ MeV. The solid line is
for the chiral model }$f_{PS,\pi}=${\protect\footnotesize \ }$f_{\pi}%
${\protect\footnotesize , and the dashed line is for the instanton model
}$f_{DP,\pi}=${\protect\footnotesize \ }$f_{\pi}${\protect\footnotesize .}}%
\label{MvsL}%
\end{figure}

Fig. \ref{MvsL} shows the parameter space where the pion decay constant is
fixed by its value taken in the chiral limit $f_{\pi}=85$ MeV
\cite{Gasser:1984gg}.

For the instanton model (\ref{QPiVertI}), the pion DA (\ref{PionDAsym}) is
reduced to%
\begin{align}
&  \varphi_{\pi}^{\mathrm{I}}\left(  x\right)  =\frac{N_{c}}{4\pi
^{2}f_{\mathrm{PS},\pi}^{2}}M_{q}\int_{0}^{\infty}\frac{dL}{L}\left(
x\sigma_{m}\left(  xL\right)  \sigma\left(  \overline{x}L\right)
+\overline{x}\sigma\left(  xL\right)  \sigma_{m}\left(  \overline{x}L\right)
\right)  ,\quad\label{PionDAsymInst}\\
&  \varphi_{\pi}^{I}\left(  x=0\right)  =0,\qquad\int_{0}^{1}dx\varphi_{\pi
}^{\mathrm{I}}\left(  x\right)  =1.\nonumber
\end{align}
For the chiral model (\ref{QPiVertT}), one obtains the pion DA%
\begin{align}
&  \varphi_{\pi}^{\chi}\left(  x\right)  =\frac{N_{c}}{8\pi^{2}f_{\pi}^{2}%
}\int_{0}^{\infty}\frac{dL}{L}\left(  xd_{m2}\left(  xL\right)  d\left(
\overline{x}L\right)  +\overline{x}d\left(  xL\right)  d_{m2}\left(
\overline{x}L\right)  +d_{m}\left(  xL\right)  d_{m}\left(  \overline
{x}L\right)  \right)  ,\label{PionDAsymChiral}\\
&  \varphi_{\pi}^{\chi}\left(  x=0\right)  =\frac{N_{c}}{4\pi^{2}f_{\pi}^{2}%
}\int_{0}^{\infty}du\frac{m^{2}\left(  u\right)  }{D\left(  u\right)  }%
,\qquad\int_{0}^{1}dx\varphi_{\pi}^{\mathrm{\chi}}\left(  x\right)
=1\label{PionDAsymChiral0}%
\end{align}
which is not vanishing at the endpoints $x=0$ and $x=1.$

In above expressions, we used the following notations for the correspondence
between momentum and $\alpha$-representation (in addition to definitions
(\ref{Dalpha}), (\ref{Falpha}))%
\begin{align}
&  \frac{m^{2}\left(  k^{2}\right)  }{D\left(  k^{2}\right)  }\sim
d_{m2}\left(  \alpha\right)  ,\nonumber\\
&  \frac{f\left(  k^{2}\right)  }{D\left(  k^{2}\right)  }\sim\sigma\left(
\alpha\right)  ,\quad\frac{m\left(  k^{2}\right)  f\left(  k^{2}\right)
}{D\left(  k^{2}\right)  }\sim\sigma_{m}\left(  \alpha\right)  .\nonumber
\end{align}
The explicit form of the functions in $\alpha$ representation in the case of
the model defined by (\ref{DynMass}) and (\ref{GaussProf}) is given in Appendix.

\begin{figure}[h]
\begin{minipage}[c]{8cm}
\includegraphics[width=1.0\textwidth]{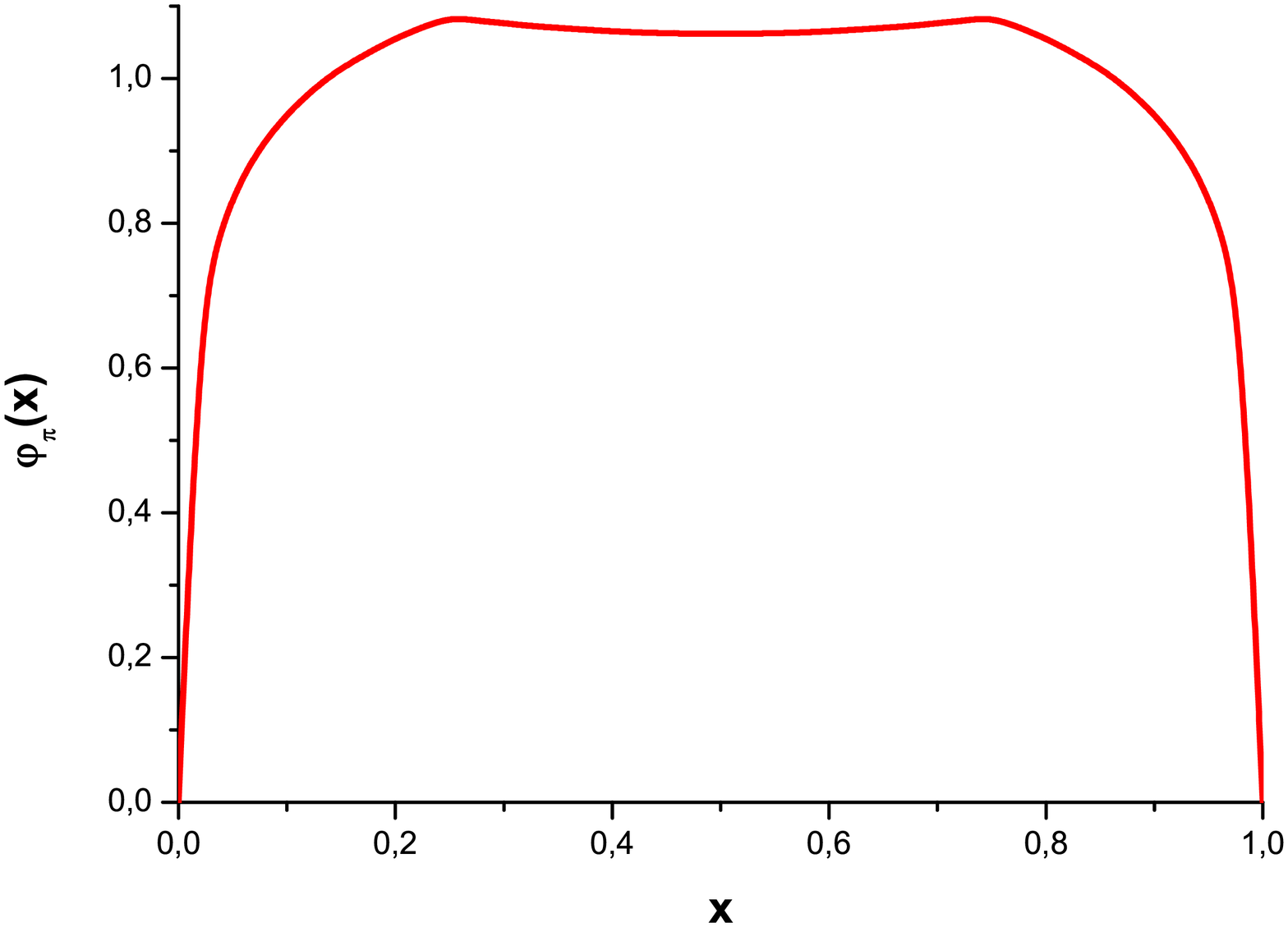}
\end{minipage}
\begin{minipage}[c]{8cm}
\includegraphics[width=1.0\textwidth]{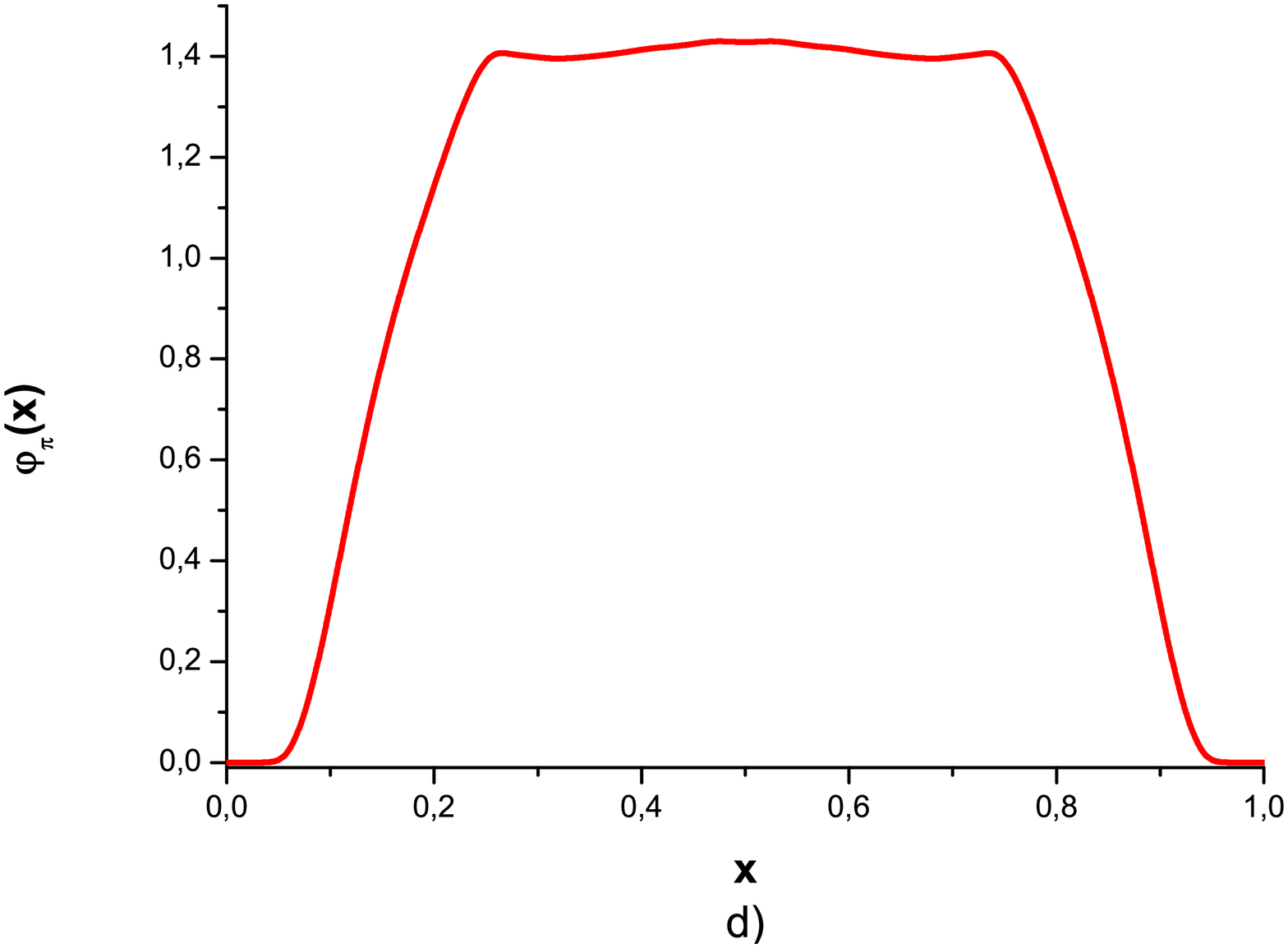}
\end{minipage}
\par
\begin{minipage}[c]{8cm}
\includegraphics[width=1.0\textwidth]{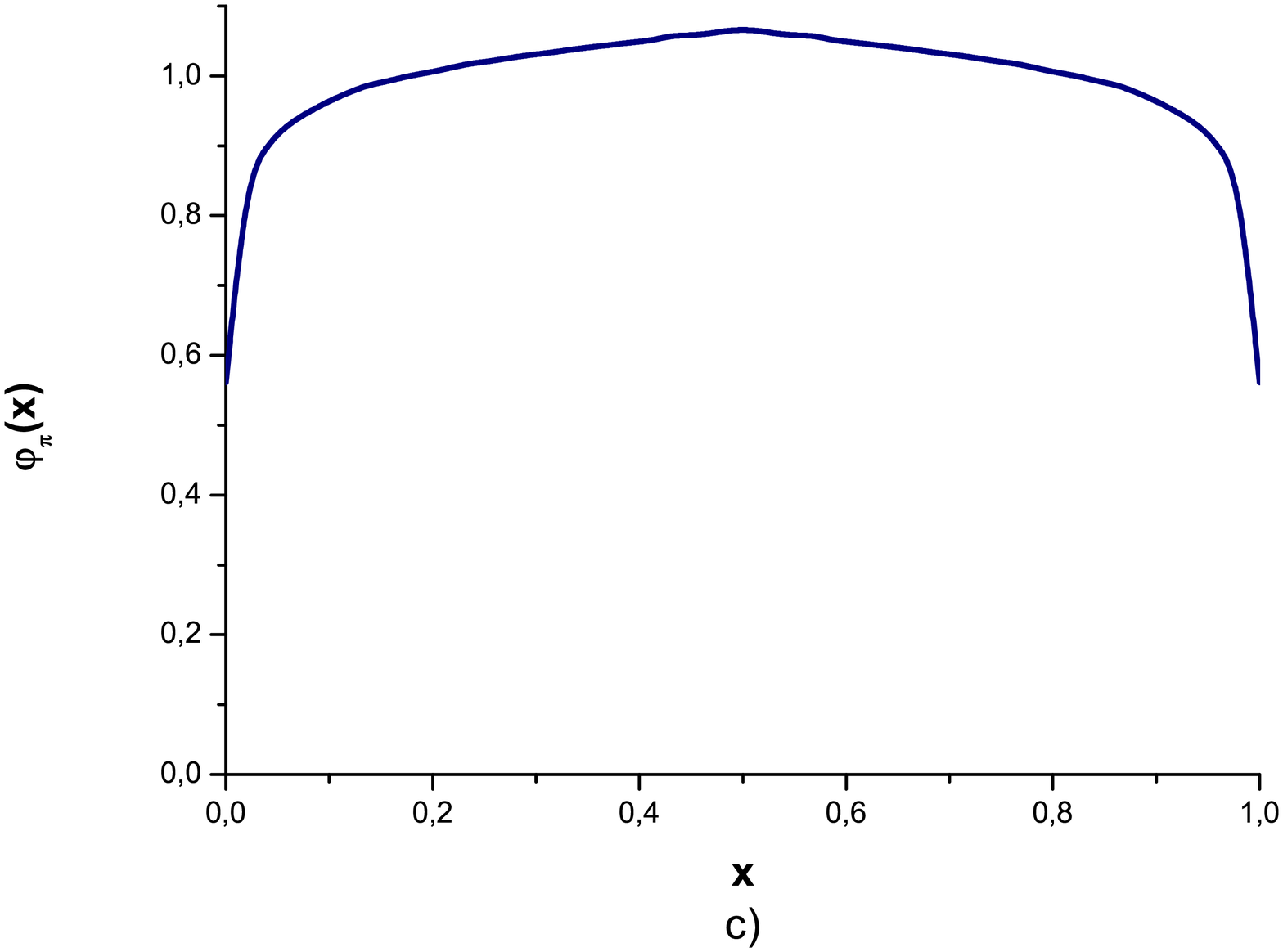}
\end{minipage}
\begin{minipage}[c]{8cm}
\includegraphics[width=1.0\textwidth]{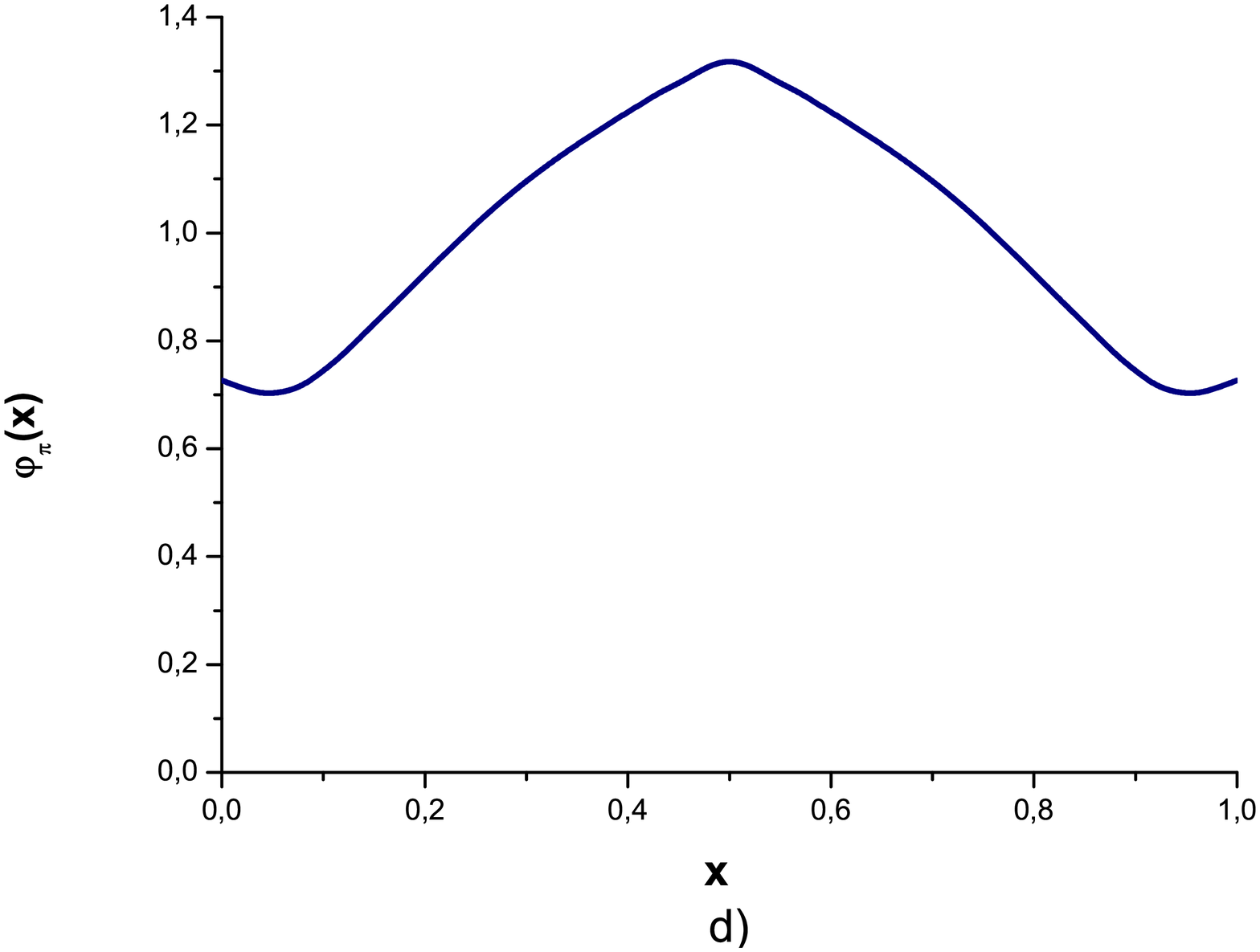}
\end{minipage}
\caption{{\protect\footnotesize Pion distribution amplitude for the instanton
model with parameters a) $M_{q}=125$ MeV, $\Lambda=0.016$ GeV$^{-2}$ and b)
$M_{q}=300$ MeV, $\Lambda=1.3$ GeV$^{-2}$; and chiral model with parameters c)
$M_{q}=125$ MeV, $\Lambda=0.0098$ GeV$^{-2}$ and d) $M_{q}=300$ MeV,
$\Lambda=0.639$ GeV$^{-2}$.}}%
\label{PiDA}%
\end{figure}

In Fig. \ref{PiDA} the different shapes of the pion DA are shown as they are
calculated within the instanton and chiral models for the values of the
dynamical quark mass $M_{q}=300$ MeV and $M_{q}=125$ MeV. The parameter
$\Lambda$ is defined to fit the pion decay constant in chiral limit $f_{\pi
}=85$ MeV. For smaller $M_{q}$ the pion DA is close to a flat shape. For
larger $M_{q}$ it is more sensitive to the nonlocal part of the photon vertex
and, in case of the instanton model, it is strongly suppressed in the vicinity
of endpoints. \begin{figure}[ph]
\hspace*{-10mm}\includegraphics[width=0.8\textwidth]{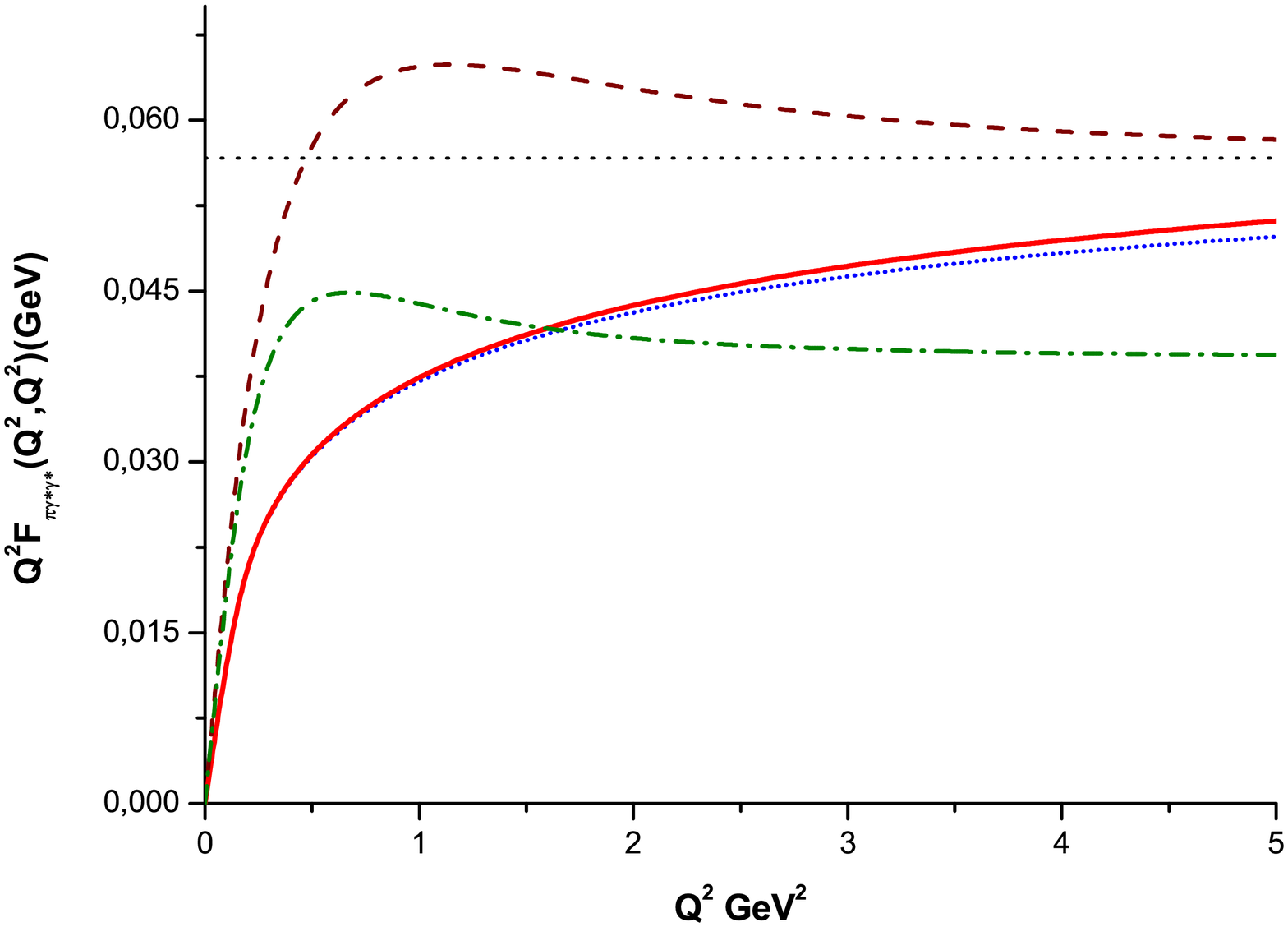} \vspace
*{-10mm}\caption{{\protect\footnotesize Photon-pion transition form factor in
symmetric kinematics for the instanton model with parameters $M_{q}=125$ MeV,
$\Lambda=0.016$ GeV$^{-2}$ (short pointed line), $M_{q}=300$ MeV,
$\Lambda=1.3$ GeV$^{-2}($dash-dotted line); and chiral model with parameters
$M_{q}=125$ MeV, $\Lambda=0.0098$ GeV$^{-2}$ (solid line) and $M_{q}=300$ MeV,
$\Lambda=0.639$ GeV$^{-2}$ (dashed line). The straight dotted line is
asymptotic limit }$2f_{\pi}/3${\protect\footnotesize .}}%
\label{PiFFsym}%
\end{figure}

In Fig. \ref{PiFFsym} the prediction for the pion transition form factor in
symmetric kinematics calculated from (\ref{Ampl})-(\ref{Mnonloc}) is
presented. The explicit expression for the instanton model is%
\begin{align}
F_{\pi\gamma^{\ast}\gamma^{\ast}}^{\mathrm{loc,I}}\left(  0;Q^{2}%
,Q^{2}\right)   &  =\frac{N_{c}M_{q}}{6\pi^{2}f_{\pi}}\int\frac{d\left(
\alpha\beta\gamma\right)  }{\Delta^{3}}e^{-\frac{1}{\Delta}\gamma\left(
\alpha+\beta\right)  Q^{2}}\sigma\left(  \beta\right)  \label{FQQI}\\
&  \cdot\left[  2\alpha\sigma_{m}\left(  \alpha\right)  d\left(
\gamma\right)  +\gamma\sigma\left(  \alpha\right)  d_{m}\left(  \gamma\right)
\right]  ,\nonumber
\end{align}
and for the chiral model is%
\begin{align}
F_{\pi\gamma^{\ast}\gamma^{\ast}}^{\mathrm{loc,\chi}}\left(  0;Q^{2}%
,Q^{2}\right)   &  =\frac{N_{c}}{6\pi^{2}f_{\pi}}\int\frac{d\left(
\alpha\beta\gamma\right)  }{\Delta^{3}}\alpha d\left(  \gamma\right)  \left[
e^{-\frac{1}{\Delta}\gamma\left(  \alpha+\beta\right)  Q^{2}}d_{m2}\left(
\alpha\right)  d\left(  \beta\right)  \right.  \label{FQQT}\\
&  \left.  +\left(  e^{-\frac{1}{\Delta}\gamma\left(  \alpha+\beta\right)
Q^{2}}+e^{-\frac{1}{\Delta}\alpha\left(  \beta+\gamma\right)  Q^{2}}\right)
d_{m}\left(  \alpha\right)  d_{m}\left(  \beta\right)  \right]  .\nonumber
\end{align}
As it is seen from Fig.\ref{PiFFsym}, the qualitative behavior of the pion
transition form factor for fixed quark mass is similar for the two different
models. For $M_{q}=300$ MeV, the combination $Q^{2}F_{\pi\gamma^{\ast}%
\gamma^{\ast}}$ rapidly turns into the asymptotic regime as expected in the
standard factorization scheme. The asymptotic limits are different for the two
models, $2f_{\pi}/3$ for the chiral model and $2f_{\mathrm{PS,}\pi}%
^{2}/3f_{\pi}$ for the instanton model. However, for smaller masses the effect
of vertex non-localities is diminished, in particular $f_{DP\mathrm{,}\pi
}\approx f_{PS,\pi}$ for the instanton model. One sees from Fig.
\ref{PiFFsym}, that for $M_{q}=125$ MeV the behavior of the form factors is
similar for both models.

\section{The BABAR data within the instanton and chiral models}

Let us consider the model predictions for the pion transition form factor in
the asymmetric kinematics ($q_{1}^{2}=Q^{2},q_{2}^{2}=0$) calculated from
(\ref{Ampl})-(\ref{Mnonloc}) in the region, where experimental data exist. The
explicit expression for the instanton model is%
\begin{align}
F_{\pi\gamma^{\ast}\gamma}^{\mathrm{loc,I}}\left(  0;Q^{2},0\right)   &
=\frac{N_{c}M_{q}}{6\pi^{2}f_{\pi}}\int\frac{d\left(  \alpha\beta
\gamma\right)  }{\Delta^{3}}e^{-\frac{\alpha\gamma}{\Delta}Q^{2}%
}\label{FIinst}\\
&  \cdot\left[  \left(  \alpha\sigma_{m}\left(  \alpha\right)  \sigma\left(
\beta\right)  +\beta\sigma\left(  \alpha\right)  \sigma_{m}\left(
\beta\right)  \right)  d\left(  \gamma\right)  +\gamma\sigma\left(
\alpha\right)  \sigma\left(  \beta\right)  d_{m}\left(  \gamma\right)
\right]  ,\nonumber
\end{align}
and for the chiral model is%
\begin{align}
F_{\pi\gamma^{\ast}\gamma}^{\mathrm{loc,\chi}}\left(  0;Q^{2},0\right)   &
=\frac{N_{c}}{12\pi^{2}f_{\pi}}\int\frac{d\left(  \alpha\beta\gamma\right)
}{\Delta^{3}}e^{-\frac{\alpha\gamma}{\Delta}Q^{2}}\left\{  \gamma d_{m}\left(
\alpha\right)  d\left(  \beta\right)  d_{m}\left(  \gamma\right)  \right.
\label{FIIchi}\\
&  \left.  +d\left(  \gamma\right)  \left[  \alpha d_{m2}\left(
\alpha\right)  d\left(  \beta\right)  +d\left(  \alpha\right)  \beta
d_{m2}\left(  \beta\right)  +\left(  2\alpha+\beta\right)  d_{m}\left(
\alpha\right)  d_{m}\left(  \beta\right)  \right]  \right\}  .\nonumber
\end{align}

\begin{figure}[h]
\hspace*{-10mm}\includegraphics[width=0.8\textwidth]{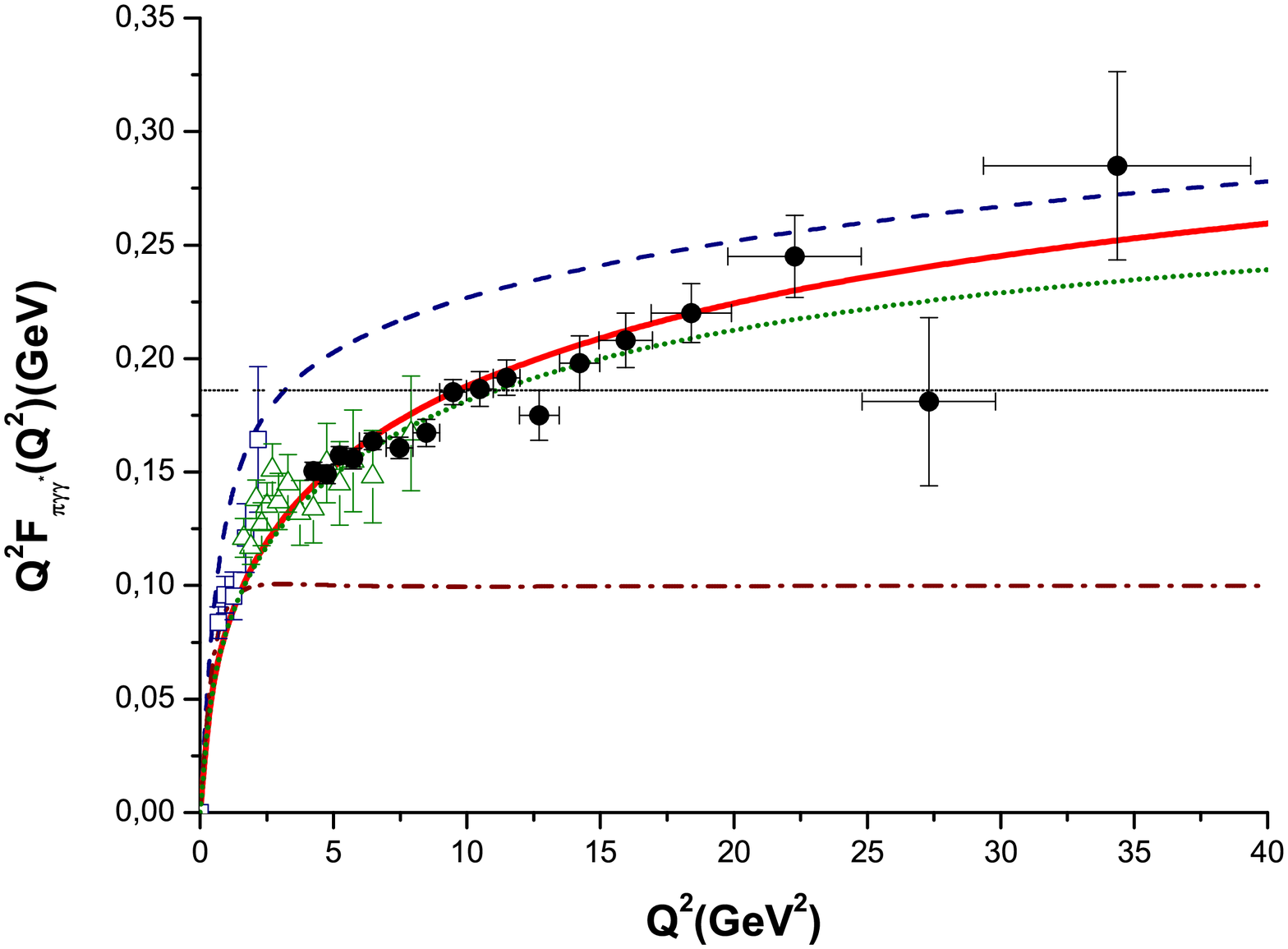}
\vspace*{-10mm}\caption{{\protect\footnotesize Photon-pion transition form
factor in asymmetric kinematics for the instanton model with parameters
$M_{q}=125$ MeV, $\Lambda=0.016$ GeV$^{-2}$ (short pointed line), $M_{q}=300$
MeV, $\Lambda=1.3$ GeV$^{-2}($dash-dotted line); and chiral model with
parameters $M_{q}=125$ MeV, $\Lambda=0.0098$ GeV$^{-2}$ (solid line) and
$M_{q}=300$ MeV, $\Lambda=0.639$ GeV$^{-2}$ (dashed line). The straight dotted
line is asymptotic limit }$2f_{\pi}${\protect\footnotesize .}}%
\label{AsymA01}%
\end{figure}

\begin{figure}[h]
\centering\hspace{-4cm} \begin{minipage}[c]{8cm}
\includegraphics[width=1.5\textwidth]{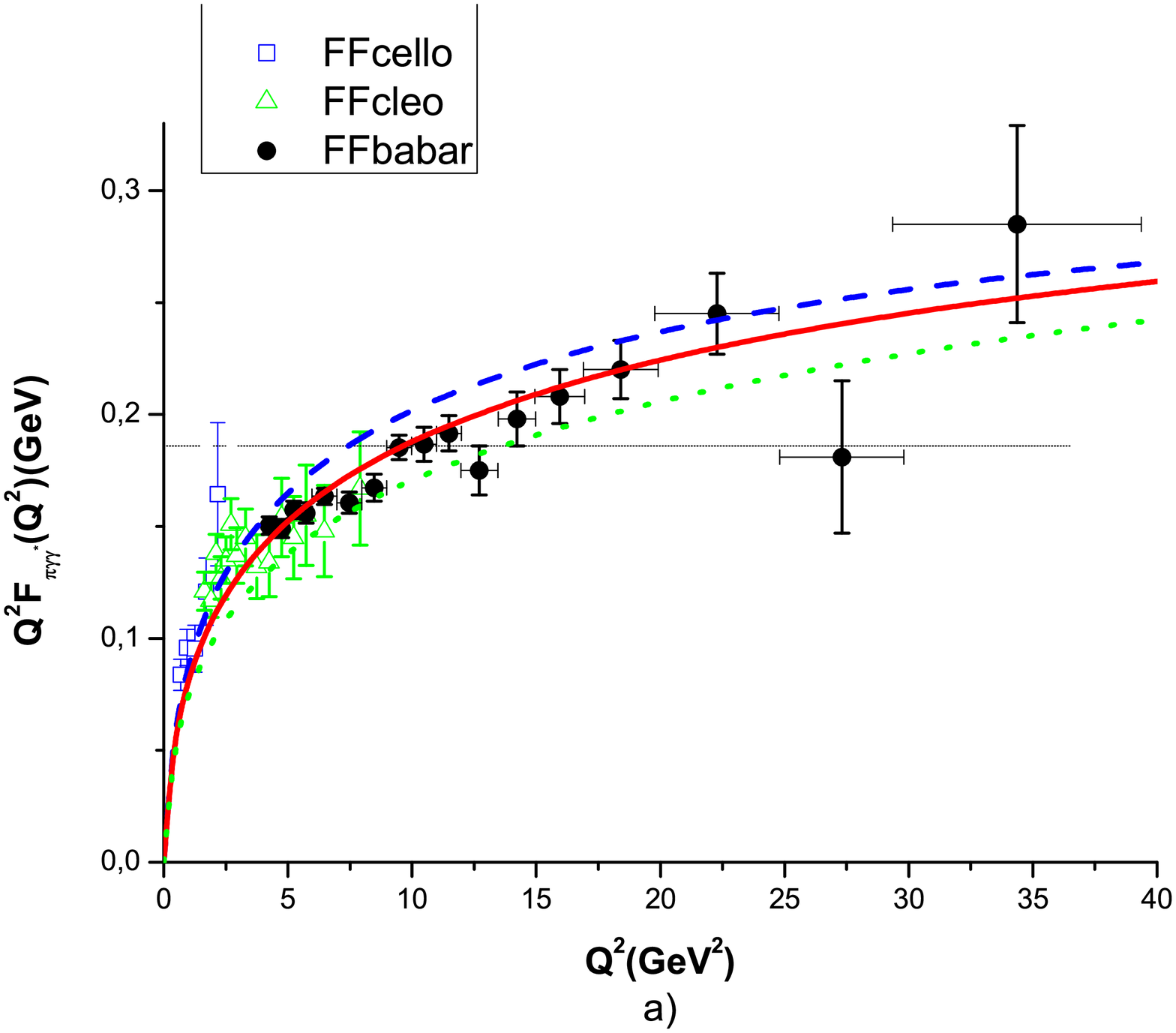}
\end{minipage}
\newline\centering\hspace{-4cm} \begin{minipage}{8cm}
\includegraphics[width=1.5\textwidth]{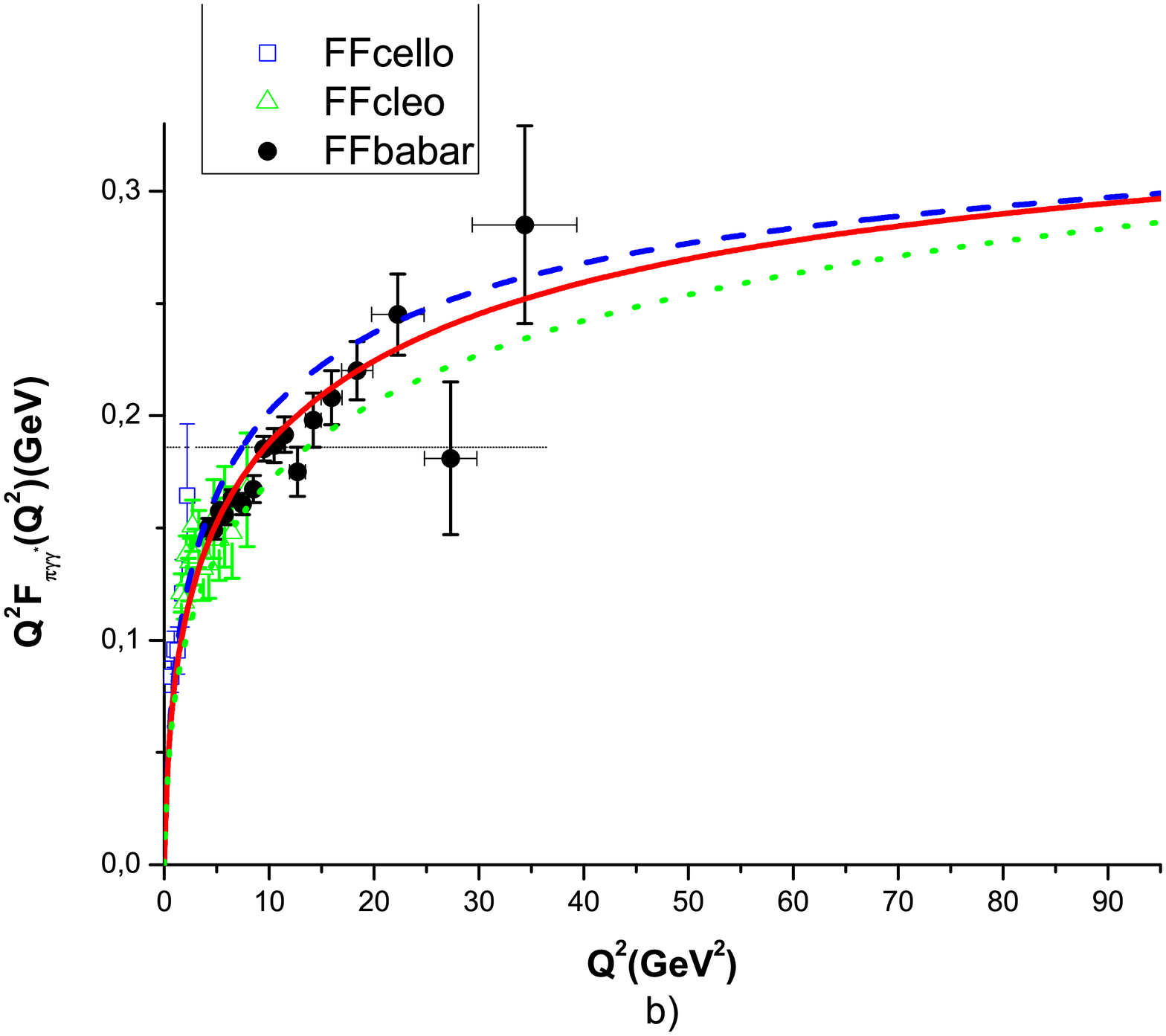}
\end{minipage}
\caption{{\protect\footnotesize Photon-pion transition form factor in
asymmetric kinematics for the chiral model with parameters $M_{q}=125$ MeV,
$\Lambda=0.0098$ GeV$^{-2}$ (solid line), $M_{q}=135$ MeV, $\Lambda=0.0203$
GeV$^{-2}$ (dashed line), $M_{q}=115$ MeV, $\Lambda=0.0038$ GeV$^{-2}$ (dotted
line). The straight dotted line is asymptotic limit }$2f_{\pi}$%
{\protect\footnotesize .}}%
\label{PiFFasymA1}%
\end{figure}

\begin{figure}[h]
\centering\hspace{-4cm} \begin{minipage}[c]{8cm}
\includegraphics[width=1.5\textwidth]{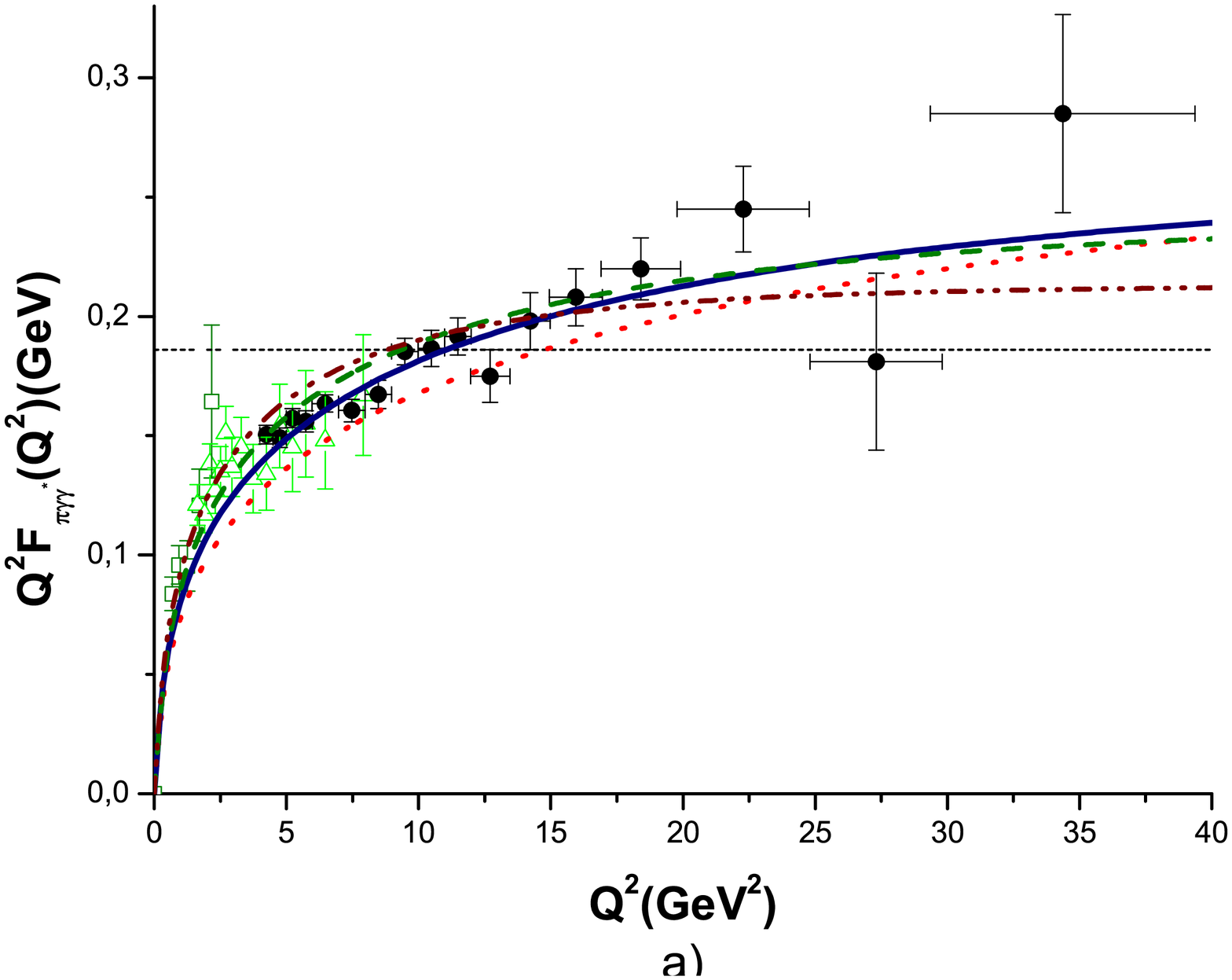}
\end{minipage}
\newline\centering\hspace{-4cm} \begin{minipage}{8cm}
\includegraphics[width=1.5\textwidth]{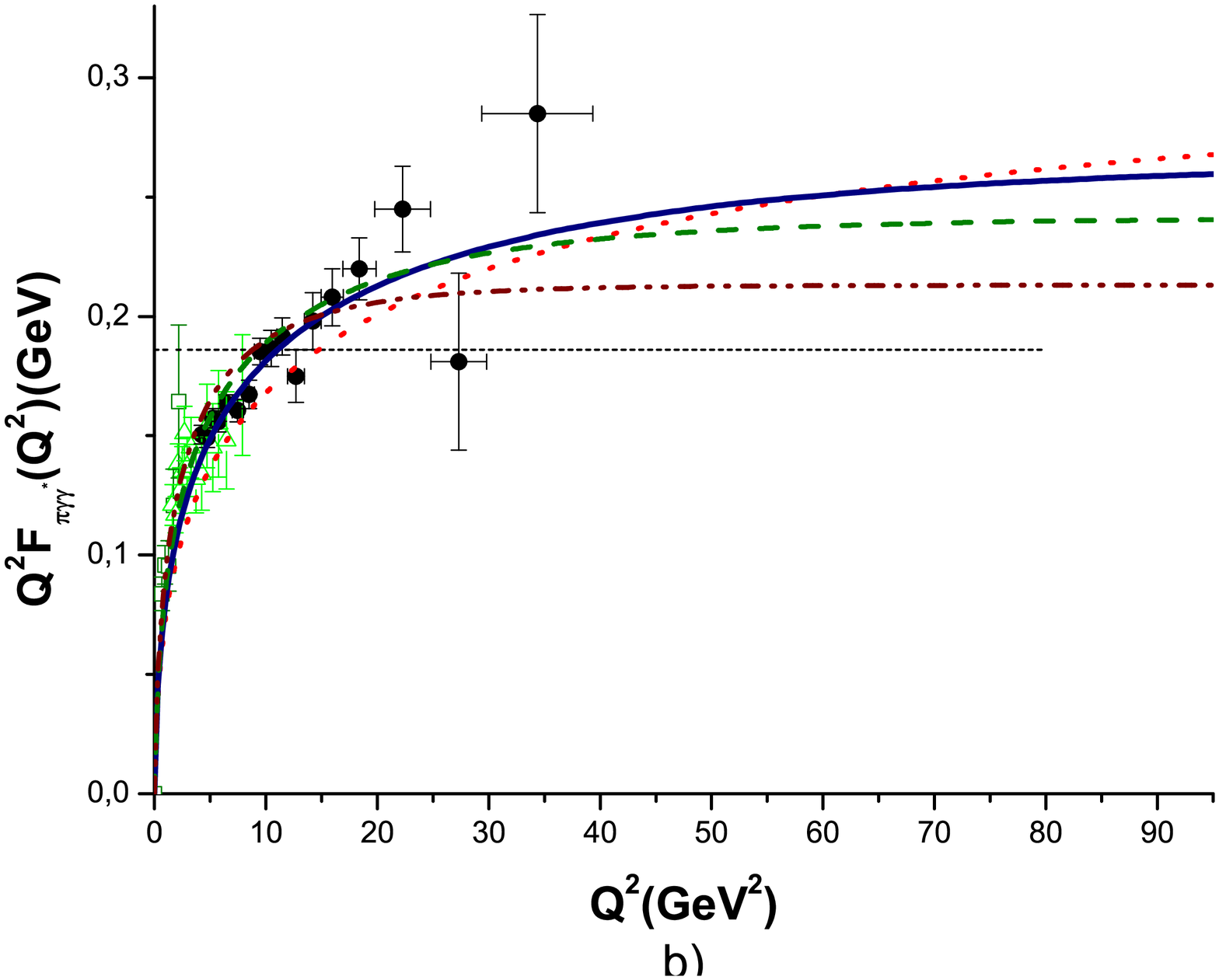}
\end{minipage}
\caption{{\protect\footnotesize Photon-pion transition form factor in
asymmetric kinematics for the instanton model with parameters $M_{q}=115$ MeV,
$\Lambda=0.0038$ GeV$^{-2}$ (dotted line), $M_{q}=125$ MeV, $\Lambda=0.0098$
GeV$^{-2}$ (solid line), $M_{q}=135$ MeV, $\Lambda=0.0203$ GeV$^{-2}$ (dashed
line), $M_{q}=150$ MeV, $\Lambda=0.077$ GeV$^{-2}$ (dash-dot-dotted line). The
straight dotted line is asymptotic limit }$2f_{\pi}${\protect\footnotesize .}}%
\label{PiFFasymA0}%
\end{figure}

In Fig. \ref{AsymA01}, we show the predictions for different values of $M_{q}%
$.For a quark mass $M_{q}=300$ MeV the model dependence is very strong and the
theoretical curves are very far from the experimental points. The chiral model
overshoots the data, while the instanton model, in correspondence with the
standard factorization scenario, shows the asymptotic $1/Q^{2}$ behavior very
early, already at $Q^{2}\sim1$ GeV$^{2}$. \ It is clearly seen, that in order
to describe the BABAR data, one has to take the dynamical quark mass
$M_{q}\approx125$ MeV. Then both models have an qualitatively good
description, with some preference to the chiral model. In Figs. \ref{PiFFasymA1}a and \ref{PiFFasymA0}a we show that the
parameter space that describes the data up to $40$ GeV$^{2}$ is rather narrow.
For the chiral model it is $M_{q}\approx125\pm10$ MeV, and for the instanton
model it is $M_{q}\approx130\pm5$ MeV. Thus in this region the instanton model
simulate the logarithmically enhanced behavior due to rather flat pion DA.
However, the further behavior of the form factor is rather different for
different models as it is seen in Figs. \ref{PiFFasymA1}b and \ref{PiFFasymA0}%
b, where the kinematical region up to $100$ GeV$^{2}$ is shown. The instanton
model finally reach its actual asymptotic $1/Q^{2}$ that follows from
(\ref{FNLac}) and (\ref{InvMom}) with the asymptotic coefficient given by%
\begin{equation}
J^{I}=\frac{N_{c}}{4\pi^{2}f_{\mathrm{PS},\pi}^{2}}M_{q}\int_{0}^{\infty
}du\frac{uf\left(  u\right)  }{D\left(  u\right)  }\int_{0}^{1}dy\frac
{f\left(  yu\right)  m\left(  yu\right)  }{D\left(  yu\right)  }\label{Jinst}%
\end{equation}
For the chiral model the logarithmic growth continues for all $Q^{2}$ with the
asymptotics following from (\ref{FIIas})%
\begin{align}
& F_{\pi\gamma^{\ast}\gamma}^{As\mathrm{,\chi}}\left(  0;Q^{2},0\right)
\overset{Q^{2}\rightarrow\infty}{=}\frac{1}{Q^{2}}\frac{N_{c}}{12\pi^{2}%
f_{\pi}}\left[  \int_{0}^{\infty}du\frac{m^{2}\left(  u\right)  }{D\left(
u\right)  }\ln\left(  \frac{Q^{2}}{u}\right)  +A^{\chi}\right]
,\label{FAsChiral}\\
& A^{\chi}=\int_{0}^{\infty}du\frac{m\left(  u\right)  }{D\left(  u\right)
}\int_{0}^{1}dy\frac{m\left(  yu\right)  }{D\left(  yu\right)  }\left[
u-2m\left(  u\right)  m\left(  yu\right)  \right]  .\nonumber
\end{align}


\begin{figure}[th]
\hspace*{-10mm}\includegraphics[width=0.8\textwidth]{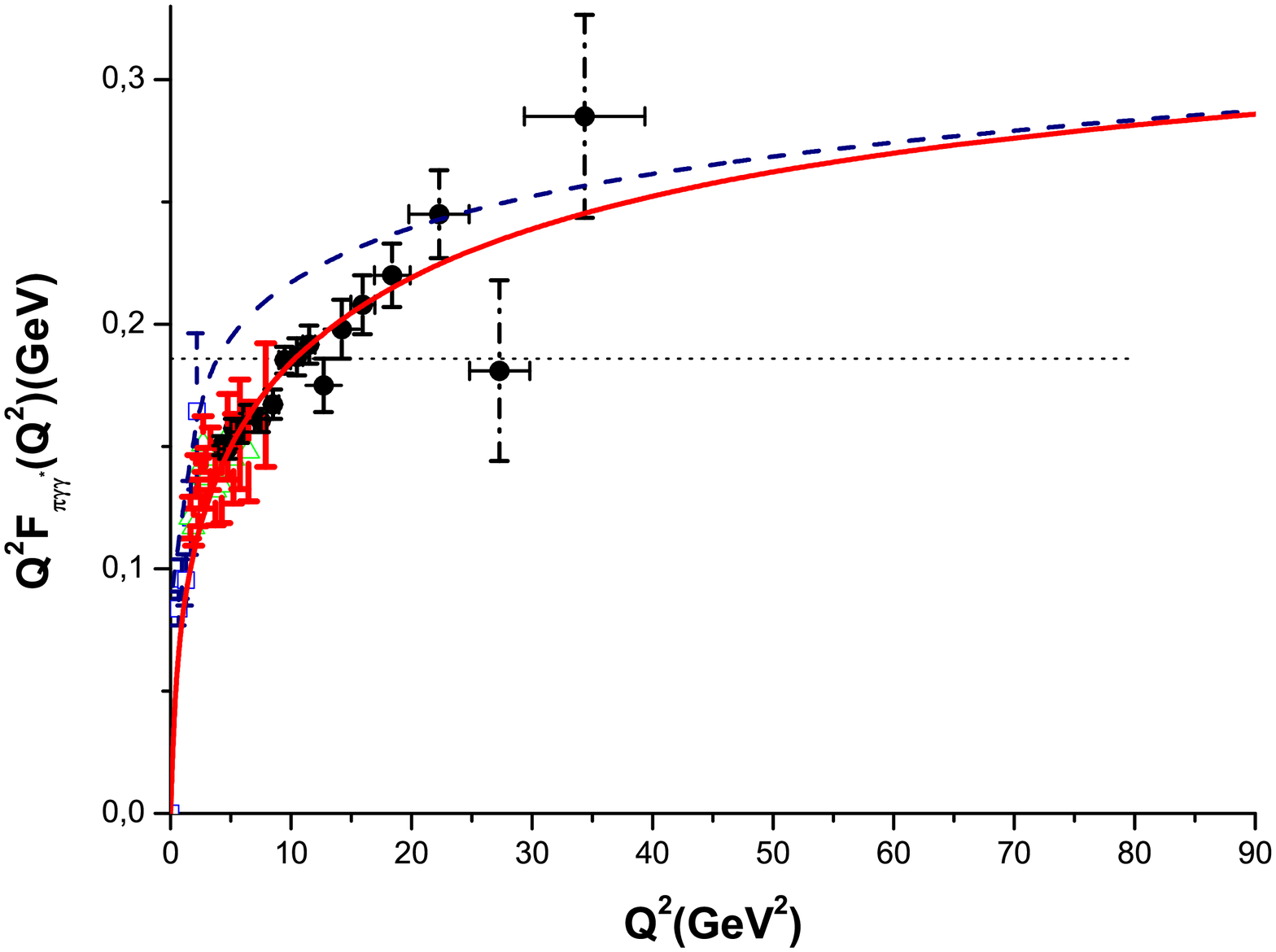}
\vspace*{-10mm}\caption{{\protect\footnotesize Photon-pion transition form
factor (solid line) and its asymptotic part (dashed line) in asymmetric
kinematics for the chiral model with parameters $M_{q}=125$ MeV,
$\Lambda=0.0098$ GeV$^{-2}$ . }}%
\label{FFasympt}%
\end{figure}

Let us make few comments. First of all, the form factor and its asymptotics
are rather different at lower $Q^{2}$. From Fig. \ref{FFasympt} it is seen
that the asymptotic curve conjugates the calculated curve in the region of
order of 100 GeV$^{2}$. Secondly, the fact, that the quark mass leading to a
satisfactory fit of the data is quite small, is not fully unexpected. There
are not many quantities that are very sensitive to the dynamical quark mass.
The precisely known contribution of the hadronic vacuum to the anomalous
magnetic moment of muon, $g-2$, is infrared sensitive and demands low values
for the quark mass, $M_{q}\approx200$ MeV
\cite{Milton:2001mq,Pivovarov:2001mw,Dorokhov:2004ze}. Finally, remember also,
that understanding the asymptotics of the pion transition form factor is
important for selection of realistic nonperturbative models, used to estimate
the hadronic contribution of the light-by-light process to $g-2$
\cite{Melnikov:2003xd,Dorokhov:2008pw}.

\section{Conclusions}

The present paper is devoted to the so-called BABAR puzzle. New very precise
data were obtained by the BABAR collaboration for the photon-pion transition
form factor in very wide kinematical region up to large photon virtualities
$Q^{2}\approx40$ GeV$^{2}$ \cite{:2009mc}. The data overshoot the asymptotic
limit for $Q^{2}F_{\pi\gamma\gamma^{\ast}}\left(  Q^{2}\right)  $ predicted by
Brodsky and Lepage \cite{Brodsky:1981rp}, and have a tendency to grow further.
Both facts are in strong contradiction with the standard QCD factorization
approach, which constitutes the BABAR puzzle.

The main problem is the unstopped growth of the new data points for
$Q^{2}F_{\pi\gamma\gamma^{\ast}}\left(  Q^{2}\right)  $ that is inconsistent
with the predicted $Q^{2}F_{\pi\gamma\gamma^{\ast}}\left(  Q^{2}\right)
\rightarrow\mathrm{constant}$, following from simple asymptotic properties of
the massless quark propagator. The key point, to solve this problem, is to
consider the properties of the pion vertex function $F(k_{1}^{2},k_{2}^{2})$
which is the analog of the light-cone pion wave function. There are two
possibilities for the momentum dependence of the pion vertex function. In the
limit, when one quark virtuality, $k_{1}^{2}$, goes to infinity, and the
other, $k_{2}^{2}$, remains finite, the vertex function may not necessarily
tend to zero. When it goes to zero, the pion DA $\varphi_{\pi}(x),$ which is a
functional of the pion vertex function, is zero at the endpoints,
$\varphi_{\pi}(0)=\varphi_{\pi}(1)=0$, with either strong or weak suppression
in the neighborhood of the endpoints $x=0$ and $x=1$. For the situation of
strong suppression, the asymptotic $1/Q^{2}$ behavior of the pion form factor
in asymmetric kinematics ($Q_{1}^{2}=Q^{2},Q_{2}^{2}=0$) is developed very
early, in contradiction with the BABAR data. For weak suppression (resembling
a flat distribution amplitude of the pion), the asymptotic $1/Q^{2}$ behavior
is developed quite late, and can give a reasonable description of the data in
the BABAR region with a $\ln Q^{2}/Q^{2}$ behavior in this region. For the
other case of non-vanishing pion vertex function in the above limit, the pion
DA $\varphi_{\pi}(x)$ is not zero at the endpoints, and therefore the
asymptotic $\ln Q^{2}/Q^{2}$ behavior persists over the whole range, in
particular in the BABAR region.

In order to fit the available data on the photon-pion transition form factor
from CELLO, CLEO and BABAR, we have analyzed the parameter space of two
examples of nonperturbative models, motivated by the instanton
\cite{Diakonov:1985eg} and the chiral \cite{Holdom:1990iq} models,
characterized by the two parameters, dynamical quark mass $M_{q}$ and the
parameter of non-locality $\Lambda$. The main conclusion is, that the fit to
the data requires a quite small dynamical quark mass $M_{q}\approx125$ MeV
with rather small uncertainty. As a consequence, the parameter of
non-locality, that fits the pion decay constant $f_{\pi}$, is very small,
$\Lambda\sim0.01$ GeV$^{-2}$. Thus, one has an almost local quark model with
very flat regulators in momentum space, that considerably diminishes the
difference between the nonperturbative models considered in this work. In this
respect, this situation resembles the fit by a simple local quark model, made
in \cite{Dorokhov:2009dg} with $M_{q}=135$ MeV. On the other hand, in
\cite{Radyushkin:2009zg} only the leading asymptotics were used and large mass
parameter of order of $1$ GeV were required, to fit the BABAR data.

Finally we would like to point out, that in the present work, we did not
consider QCD evolution. In \cite{Radyushkin:2009zg}, it was argumented, that
the flat pion DA corresponds to a very small momentum scale, and hence QCD
evolution is frozen. Our calculations support this point of view. In
particular, our choice of parameters fitting the BABAR data leads to quite
large values of the quark condensate, also corresponding to a very low
normalization point.

Concluding we may say, that the BABAR data being unique in their accuracy and
covering a very wide kinematical range, are consistent with considerations
based on nonperturbative QCD\ dynamics and may indicate specific properties of
the pion wave function.

\section{Acknowledgments}

The author especially thanks S.V. Mikhailov and A.V. Radyushkin, and also W.
Broniowski, S.B. Gerasimov, S.I. Eidelman, M.A. Ivanov, N.I. Kochelev, E.A.
Kuraev, H.-P. Pavel, A.A. Pivovarov, A.E. Radzhabov for discussions on the
interpretation of the high momentum transfer BABAR data for the pseudoscalar
meson transition form factors. The author acknowledges partial support from
the Scientific School grant 4476.2006.2 and the Russian Foundation for Basic
Research projects no. 10-02-00368.

\section{Appendix}

Here, the explicit expressions of the functions in $\alpha$ representation for
the Gaussian model defined by (\ref{DynMass}) and (\ref{GaussProf}) are given
\begin{align*}
d\left(  \alpha\right)    & =1+\sum_{n=1}^{\infty}\frac{\left(  -1\right)
^{n}}{n!}\left[  M_{q}^{2}\left(  \alpha-4\Lambda n\right)  \right]
^{n}\Theta\left(  \alpha-4\Lambda n\right)  ,\\
d_{m}\left(  \alpha\right)    & =M_{q}\sum_{n=0}^{\infty}\frac{\left(
-1\right)  ^{n}}{n!}\left[  M_{q}^{2}\left(  \alpha-\Lambda\left(
2+4n\right)  \right)  \right]  ^{n}\Theta\left(  \alpha-\Lambda\left(
2+4n\right)  \right)  ,\\
d_{m2}\left(  \alpha\right)    & =M_{q}^{2}\sum_{n=0}^{\infty}\frac{\left(
-1\right)  ^{n}}{n!}\left[  M_{q}^{2}\left(  \alpha-\Lambda\left(
4+4n\right)  \right)  \right]  ^{n}\Theta\left(  \alpha-\Lambda\left(
4+4n\right)  \right)  ,\\
\sigma\left(  \alpha\right)    & =\sum_{n=0}^{\infty}\frac{\left(  -1\right)
^{n}}{n!}\left[  M_{q}^{2}\left(  \alpha-\Lambda\left(  1+4n\right)  \right)
\right]  ^{n}\Theta\left(  \alpha-\Lambda\left(  1+4n\right)  \right)  ,\\
\sigma_{m}\left(  \alpha\right)    & =M_{q}\sum_{n=0}^{\infty}\frac{\left(
-1\right)  ^{n}}{n!}\left[  M_{q}^{2}\left(  \alpha-\Lambda\left(
3+4n\right)  \right)  \right]  ^{n}\Theta\left(  \alpha-\Lambda\left(
3+4n\right)  \right)  .
\end{align*}


\end{document}